\documentclass[a4paper,10pt]{article}
\usepackage{a4wide}
\usepackage[utf8]{inputenc}
\usepackage[T1]{fontenc}

\usepackage{subcaption}
\usepackage{amsmath,amssymb}
\usepackage{amsthm}
\usepackage{enumerate} 
\usepackage{cancel}

\usepackage{authblk}

\usepackage[round]{natbib}
\bibpunct[, ]{(}{)}{;}{a}{}{,}

\usepackage[svgnames]{xcolor}
\usepackage[pdftex]{hyperref}
\hypersetup{
    colorlinks=true,
    colorlinks,
    linkcolor=Navy,
    citecolor=Navy,
    urlcolor=Navy
}
\definecolor{highlightNEW}{named}{black}

\usepackage{graphicx}
\graphicspath{{fig/}}
\usepackage{epstopdf}
\usepackage{float}
\usepackage{tabularx}
\usepackage{booktabs}
\usepackage{multirow}
\usepackage{arydshln}

\newdimen\CdotAxis
\newcommand*{\CdotAux}[3]{%
  {%
    \settoheight\CdotAxis{$#2\vcenter{}$}%
    \sbox0{%
      \raisebox\CdotAxis{%
        \scalebox{#1}{%
          \raisebox{-\CdotAxis}{%
            $\mathsurround=0pt #2#3$%
          }%
        }%
      }%
    }%
    \dp0=0pt %
    \sbox2{$#2\bullet$}%
    \ifdim\ht2<\ht0 %
      \ht0=\ht2 %
    \fi
    \sbox2{$\mathsurround=0pt #2#3$}%
    \hbox to \wd2{\hss\usebox{0}\hss}%
  }%
}
\makeatletter
\def\mathcolor#1#{\@mathcolor{#1}}
\def\@mathcolor#1#2#3{%
  \protect\leavevmode
  \begingroup
    \color#1{#2}#3%
  \endgroup
}
\makeatother

\let\oldalpha\alpha
\renewcommand{\alpha}{\mathcolor{highlightNEW}{\oldalpha}}

\newcommand{\email}[1]{\href{mailto:#1}{#1}}

\newcommand\blfootnote[1]{%
  \begingroup
  \renewcommand\thefootnote{}\footnote{#1}%
  \addtocounter{footnote}{-1}%
  \endgroup
}

\title{\textcolor{Navy}{\textsc{A Distributional Framework for Behavioral Characterization and Risk Management of Large Language Models}}}

\author[1]{David Garcia-Lorite}
\author[2]{Ra\'{u}l Merino\thanks{Corresponding author, \email{raul.merino85@gmail.com}}}

\affil[1]{BBVA, S.A. - GMRU-COE, \authorcr C/ Sauceda, 28, 28050 Madrid, Spain}
\affil[2]{VidaCaixa S.A., Quantitative Risk Modeling, \authorcr C/Juan Gris, 2-8, 08014 Barcelona, Spain.}

\date{\normalfont\small\today}

\begin{document}

\maketitle

\begin{abstract}
Large language models are stochastic by nature. Repeated executions under the same experimental conditions may produce different responses. We propose a framework that can be applied across the model lifecycle, from task design to residual behavior and continuous monitoring. The method consists of building an empirical response distribution estimated through Monte Carlo simulation. Semantic interventions are introduced to study how these distributions change. We illustrate the framework with two different tasks, borrower classification and music album recommendation, and quantify the resulting distributional changes. The framework is linked to model risk management and governance by measuring robustness and control effectiveness, and characterizing residual behavior. Since it relies only on observed responses, it can also be applied to black-box models accessed through commercial APIs.
\end{abstract}

\blfootnote{
We would like to thank Wayne B. Shoumaker for suggesting that the approach in \cite{Merino2026} could be used to quantify residual model risk, and Agus Sudjianto for reading the manuscript and providing valuable insights.}

%\keywords{V--}
%\ccode{MSC classification}{--}
%\ccode{JEL classification}{--}

%\TODO{Remove ToC in final version}
%\tableofcontents
%\clearpage

\section{Introduction}

A classical problem in science is understanding how a model behaves when its inputs change. From celestial mechanics and chemistry to economics and civil engineering, mathematical models have been used to design systems, improve efficiency, reduce costs, and support decision-making. For this reason, it is important not only to understand a model under reference conditions, but also to study how its behavior changes when initial conditions, parameters, or external conditions are modified. Sensitivity analysis, stress testing, and uncertainty analysis provide different tools to characterize these changes.

As models become more relevant for organizations and society, their validation and governance tend to become formalized. In finance, for example, \cite{SR117} formalized supervisory expectations for Model Risk Management throughout the model lifecycle. Later, it was replaced by \cite{SR262}. These principles have become a reference for controlling the risks arising from the use of quantitative models.

Large language models are no different in this respect, just more complex. Traditional modeling has valued parsimony and explicit structure, making models easier to understand and validate. Large language models represent almost the opposite situation. Their behavior emerges from highly complex interactions and can depend on large amounts of contextual and semantic information. Yet the same questions remain relevant: how does the model behave under reference conditions, how sensitive is it to changes in its inputs, and what happens under alternative or stressed scenarios?

As LLMs become more widely used, governance frameworks are gradually taking shape. The \cite{EUAIAct2024} establishes regulatory requirements, while the \cite{NISTAIRMF2023} provides a voluntary framework for managing AI risks. The rapid evolution of these models makes stable validation standards difficult to define. For example, \cite{SR262} excludes generative and agentic AI models from its scope due to their novel and rapidly evolving nature. These validation challenges are also receiving attention from practitioners, as illustrated by recent work on AI validation in \cite{Sudjianto26}.

Inspired by the initial experiment presented in \cite{Merino2026}, we propose a framework for characterizing the behavior of large language models. A model response is treated as a realization of a random variable. Through repeated sampling, we estimate the empirical distribution of responses. Later, we introduce semantic interventions and study how that distribution changes. These interventions may represent close variations of the original task, demographic information, plausible user interactions, or stress scenarios. Formulating the problem as a Monte Carlo experiment offers us a broad range of statistical tools. Each empirical response distribution can be viewed as a point in the probability simplex, and differences between distributions can be quantified using distributional measures.

Recent studies have examined prompt sensitivity and response variability. For example, \cite{Chatterjee2024}, \cite{Zhuo2024}, \cite{Hua2025}, and \cite{LiuChu2026} study prompt sensitivity from different perspectives, analyzing how model outputs, scores, or internal representations change across semantically equivalent prompt formulations. Other studies focus on stochastic variability through repeated sampling. \cite{WadiFredette2025}, \cite{Rauba2025}, \cite{Manousaridis2026}, and \cite{WatanabeSun2026} use repeated model responses. \cite{WadiFredette2025} summarize repeated responses through scalar behavioral quantities and sampling statistics. \cite{Rauba2025} construct empirical distributions in a semantic-similarity space for hypothesis testing under interventions. \cite{Manousaridis2026} reduce responses to correct/incorrect outcomes within a hierarchical Beta-Binomial framework. \cite{WatanabeSun2026} compare empirical cumulative distributions of similarity scores across different agent configurations. The common idea is to study model behavior through distributions and then examine how those distributions change under interventions or alternative conditions. However, the approaches differ in what they analyze and how those changes are evaluated.

Other work focuses on the problem from a model risk and governance perspective. \cite{Dixon2026} treats the LLM as a probabilistic semantic observation model whose outputs are regularized through Bayesian filtering and propagated into uncertainty and risk measures. \cite{Sudjianto24} use distributional distances to quantify distribution shifts and support threshold-based ongoing monitoring. \cite{Szpruch2026}, in turn, approach the problem from a runtime-governance perspective, emphasizing controls and continuous monitoring of deployed systems.

This paper offers a framework to understand a model's behavior under a given task and its possible variations. We study response distributions from a geometric perspective, representing them as points in the probability simplex and measuring changes using distributional distances. This provides a complementary perspective to statistical inference approaches, such as the hypothesis testing methodology of \cite{Rauba2025}. The proposed framework is well-suited for model risk applications such as control evaluation, residual behavior characterization, and continuous validation and monitoring. More generally, the baseline distribution may be constructed using this methodology or specified externally.

The following figure summarizes the proposed methodology.
\begin{figure}[H]
    \centering
    \includegraphics[width=0.8\textwidth]{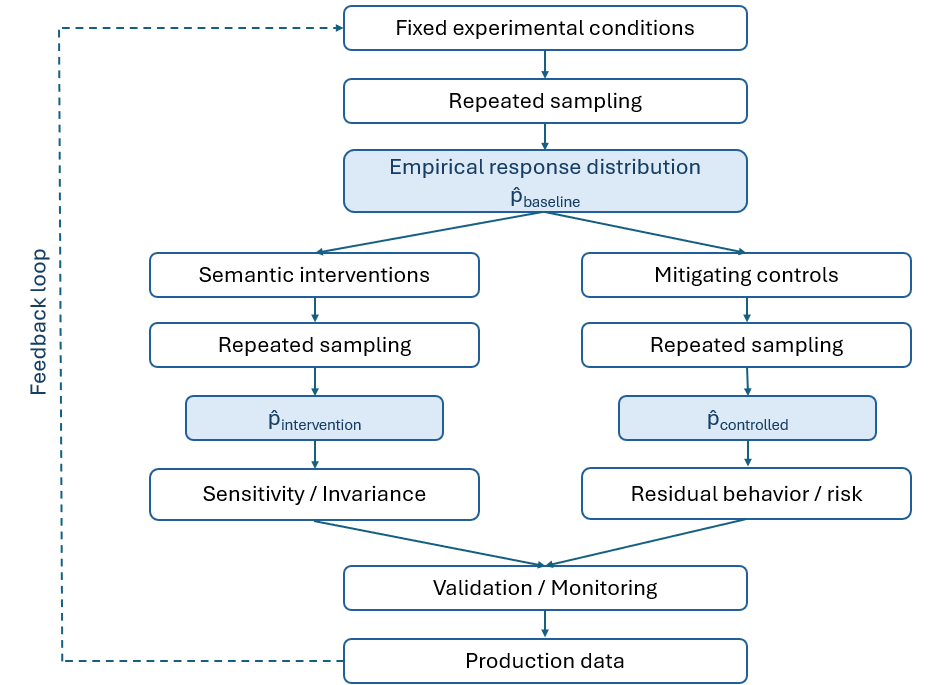}
    \caption{Proposed framework to characterize model behavior.}
    \label{fig:framework}
\end{figure}

The paper is organized as follows. Section~2 introduces the framework, including the methodology and distributional measures. Section~3 discusses its application to risk management and governance. Section~4 presents two experimental case studies. Section~5 concludes.

\section{Monte Carlo Characterization}

\subsection{Methodology}
In recent years, the use of natural language models has expanded to a wide range of applications. They answer questions, generate text, support decision-making, and act as components of larger AI systems. However, model responses are often considered individually, as if they were deterministic, even though the underlying generation process is stochastic.

For a given prompt, the model does not generate a single deterministic answer. Instead, it generates each token by sampling from a conditional probability distribution defined by the model, the prompt, and the previously generated tokens. Therefore, multiple responses may be produced under the same conditions.

Due to this construction, task conditions influence the generation process and the likelihood of possible responses. However, the model may still produce a response that does not satisfy the conditions. In some cases, due to the semantic nature of a condition, the model may interpret it differently from what we intended.

A single response is one realization of a random variable. It is insufficient to characterize the model's behavior or the range of possible responses. To characterize this behavior, we propose a Monte Carlo methodology that estimates the response distribution under fixed experimental conditions.

The methodology consists of the following steps:
\begin{enumerate}
\item Define the task to be analyzed.
\item Select the language model.
\item Fix the prompt, decoding parameters, and other relevant experimental conditions.
\item Execute the task $N$ times through the model API. Each execution is performed using a new client instance with no previous context.
\item Collect all generated responses.
\item Estimate the empirical distribution of the resulting responses.
\end{enumerate}

This Monte Carlo formulation allows standard statistical tools to be applied to the resulting response distributions.

The model may provide semantically equivalent answers written differently. Depending on the objective of the analysis, these differences can be preserved or grouped. When the interest lies in the semantic outcome rather than its linguistic realization, equivalent responses can be mapped to canonical responses to avoid an artificial fragmentation of the empirical distribution. This process can be performed manually or using automated methods, such as rules, clustering, or LLM-based semantic evaluation. In either case, the mapping procedure forms part of the measurement framework and should be validated. In this paper, responses were manually mapped to canonical outcomes.

The resulting empirical distribution provides a quantitative representation of the model's behavior. For example, it allows the characterization of response variability and the estimation of the empirical probability of each observed outcome. Since the method only requires access to generated responses, it can also be applied to black-box models accessed through external APIs. From a geometrical perspective, each empirical response distribution can be viewed as a point in the probability simplex over the observed canonical outcomes, providing a convenient representation for comparing model behavior across experimental conditions.

\subsection{Distribution Sensitivity through Semantic Interventions}

Sensitivity analysis is an essential tool in quantitative fields. In mathematics, derivatives characterize how a small perturbation in the input is translated into a change in the function's output. They provide a measure of a function's local sensitivity. Moreover, when the derivatives are bounded, they can provide information about the function's local stability.

When we validate a model, we are also interested in its behavior beyond local perturbations. This includes behavior over a wider range of inputs or under stress scenarios. These analyses provide information about the model's behavior, risks, and limitations before deployment.

The same principles are relevant to large language models. Here, the problem is more complex. Due to the nature of their inputs, there is no unique way to define distance or what constitutes a small perturbation. However, several proxies exist to measure semantic distance between prompts. In this setting, we propose to characterize the model through semantic interventions and evaluate how the resulting response distribution changes.

Semantic interventions can be classified into different categories, such as: 
\begin{enumerate}
    \item \textbf{Close variations:} Semantically close variations of the original prompt.
    \item \textbf{Plausible interactions:} Alternative prompts representing plausible user interactions.
    \item \textbf{Demographic interactions:} Adding demographic information about the user to assess whether there is a shift in the distribution of responses.
    \item \textbf{Stress scenarios:} Prompts representing stress scenarios.
\end{enumerate}

Semantic interventions can be used to test behavioral invariance. A model can be considered stable when close variations do not substantially change the resulting response distribution. When an attribute should not affect the task, a material change in the response distribution may suggest a potential bias.

In this framework, bias refers to a difference in the response distribution across conditions, while discrimination refers to a difference that results in an unfavorable outcome for a specific group. When a change is observed, it is important to examine the results to identify which responses drive it and whether it remains stable across repeated samples or similar scenarios.

A common example involves demographic information. Demographic interactions can be used to assess whether these attributes affect the response distribution. A difference does not necessarily imply discrimination. In some applications, demographic information can predict a relevant outcome.

The framework can be viewed as a mapping from experimental conditions to empirical response distributions. Sensitivity analysis studies how this mapping changes under related or alternative conditions, while invariance corresponds to cases in which the resulting distributions remain approximately stable.

\subsection{Distribution Measures}
The central object of study in this methodology is the response distribution and how it changes across experimental conditions. For this reason, we define the measures used to characterize and compare these distributions.

The basic quantity is the empirical probability. Given $N$ realizations, let $n_i$ represent the number of times response $r_i$ is observed. The empirical probability is calculated as follows:

\begin{equation*}
\hat{p}_i = \frac{n_i}{N}.
\end{equation*}

The size of the observed support is denoted by $K_{\mathrm{obs}}$. It provides a measure of response diversity. Since it is estimated from a finite number of realizations, it depends on the sample size.

The empirical response distribution can be represented by the probability vector:
\begin{equation*}
\hat{\mathbf{p}}=(\hat{p}_1,\ldots,\hat{p}_{K_{\mathrm{obs}}}),
\qquad
\sum_{i=1}^{K_{\mathrm{obs}}}\hat{p}_i=1.
\end{equation*}

The probability vector defines a point in the probability simplex. Comparing model behavior across experimental conditions corresponds to comparing points in this space.

The leading response is the one with the highest empirical probability. A high probability for the leading response indicates greater concentration, while a lower probability suggests that the probability mass is more evenly distributed.

A complementary measure is the entropy of the empirical distribution:
\begin{equation*}
H(\hat{\mathbf{p}})=-\sum_{i=1}^{K_{\mathrm{obs}}}\hat{p}_i\log_2\hat{p}_i.
\end{equation*}

Entropy characterizes the overall dispersion of probability mass across the observed responses. Lower entropy indicates greater concentration, whereas higher entropy indicates greater dispersion. The effective number of response categories is given by $2^{H(\hat{\mathbf{p}})}$.

We can also compare empirical distributions directly through their probability vectors:
\begin{equation*}
\hat{\mathbf{p}}=(\hat{p}_1,\ldots,\hat{p}_K),
\qquad
\hat{\mathbf{q}}=(\hat{q}_1,\ldots,\hat{q}_K),
\end{equation*}

Both distributions must be defined over a common support of $K$ outcomes. When the observed supports differ, the common empirical support is defined as the union of their observed supports, assigning zero probability to outcomes not observed in one of the distributions.

Several measures can be used to compare distributions within the probability simplex. We consider two complementary choices. Total variation distance has a direct probabilistic interpretation, while Jensen--Shannon divergence provides an information-theoretic measure of separation.

The total variation distance between two empirical distributions is defined as

\begin{equation*}
D_{\mathrm{TV}}(\hat{\mathbf{p}},\hat{\mathbf{q}})
=
\frac{1}{2}\sum_{i=1}^{K}|\hat{p}_i-\hat{q}_i|.
\end{equation*}

It ranges from zero for identical distributions to one for distributions with disjoint supports.

We can also compare two distributions using the Jensen--Shannon divergence. It provides a symmetric and finite measure. For two empirical distributions $\hat{\mathbf{p}}$ and $\hat{\mathbf{q}}$, let $\mathbf{m}=(\hat{\mathbf{p}}+\hat{\mathbf{q}})/2$ denote their midpoint distribution. The Kullback--Leibler divergence between two discrete probability distributions $\mathbf{a}$ and $\mathbf{b}$ over the same $K$ outcomes is

\begin{equation*}
D_{\mathrm{KL}}(\mathbf{a}\|\mathbf{b})=\sum_{i=1}^{K}a_i\log_2\left(\frac{a_i}{b_i}\right),
\end{equation*}

where terms with $a_i=0$ are defined to contribute zero. The Jensen--Shannon divergence is

\begin{equation*}
D_{\mathrm{JS}}(\hat{\mathbf{p}},\hat{\mathbf{q}})
=
\frac{1}{2}
D_{\mathrm{KL}}(\hat{\mathbf{p}}\|\mathbf{m})
+
\frac{1}{2}
D_{\mathrm{KL}}(\hat{\mathbf{q}}\|\mathbf{m}).
\end{equation*}

Using logarithms in base two:

\begin{equation*}
0
\leq
D_{\mathrm{JS}}(\hat{\mathbf{p}},\hat{\mathbf{q}})
\leq
1.
\end{equation*}

A value of zero indicates identical distributions, while a value of one corresponds to the maximum divergence, which occurs when the two distributions have disjoint supports. Thus, the Jensen–Shannon divergence provides a scalar measure of the magnitude of distributional change, complementing the direct comparison of individual probability changes. It describes behavioral change but does not, by itself, determine whether that change is desirable or undesirable.

In this paper, the analysis uses the baseline distribution as the reference. However, the framework is more general. The same comparison measures can be applied to any benchmark distribution. This benchmark can be an empirical baseline, an externally defined or expected distribution $\mathbf{p}^\star$, or a deterministic benchmark. The deterministic benchmark can be represented as a degenerate distribution with all probability mass assigned to a single outcome. When this outcome represents the correct response, it corresponds to the usual ground-truth case.

Some applications may identify a subset of outcomes as undesirable or unsuccessful. Let $\mathcal{U}$ denote the set of such outcomes. The empirical probability mass assigned to undesirable outcomes is

\begin{equation*}
P_{\mathcal{U}}(\hat{\mathbf{p}})=\sum_{i:r_i\in\mathcal{U}}\hat{p}_i.
\end{equation*}

When meaningful costs or severities can be assigned to responses, the empirical distribution can be used to construct a severity-weighted measure. Let $C_i$ denote the severity or cost associated with response $r_i$. The expected loss is

\begin{equation*}
L(\hat{\mathbf{p}})=\sum_{i=1}^{K_{\mathrm{obs}}}\hat{p}_i C_i.
\end{equation*}

Identifying undesirable outcomes allows their probability mass to be summarized by a scalar value. Assigning severities or costs further allows the empirical distribution to be used to construct a risk measure.

All quantities defined in this section are estimated from finite Monte Carlo samples and are therefore subject to sampling uncertainty. This uncertainty can be quantified using multinomial sampling distributions or bootstrap resampling, depending on the measure. Section~4.6 illustrates the bootstrap approach for empirical response probabilities and total variation distance. Formal hypothesis testing of distributional changes can also be performed, as in \cite{Rauba2025}.

\subsection{Limitations}

The proposed methodology for characterizing the response distribution of large language models relies on standard Monte Carlo techniques. However, it has limitations.

The most obvious limitation is finite sampling. The number of iterations required depends on the desired level of precision and the underlying distribution. Tail events may remain unobserved. The methodology also treats repeated executions using separate client instances and no shared context as independent realizations. In a statistical sense, independence cannot be verified when using closed models through external APIs.

Another limitation is its cost. The method requires repeated model executions for the baseline and for each selected intervention. The required API calls incur computational and monetary costs. This becomes particularly relevant when scaling and deploying models. 

One limitation arises from the nature of the input space. The space of semantically similar prompts and possible interventions is large, and not all relevant interventions may be anticipated before deployment. Moreover, there is no universal method for measuring semantic distance between prompts. Mapping semantically equivalent responses to canonical outcomes introduces an additional source of uncertainty. Sensitivity to the mapping procedure can therefore be treated as an additional validation dimension.

Finally, the empirical characterization is conditional on the model, prompt, and experimental configuration. Similar architectures or behavioral patterns across models do not provide sufficient evidence for generalization, which requires additional empirical validation.

\section{Application to Risk Management and Governance}

Risks associated with large language models arise across sectors and applications. Any organization deploying these models may be exposed to risks related to their behavior and use.

These models introduce additional complexity, but the principles of risk management remain applicable. From this perspective, Model Risk Management frameworks in financial institutions provide a useful reference for implementing these principles in practice. 

We consider the application of the proposed methodology across the model lifecycle.

\subsection{Pre-production}
Before deploying a model, it is important to understand its behavior, use, and potential risks. The evaluation of the baseline response distribution and the changes under semantic interventions establish a framework for this assessment.

The baseline scenario provides information about the distribution under the reference conditions. We can analyze the raw responses or categorize them by meaning. We can measure response variability, identify patterns, and estimate the probability of each response. When undesirable outcomes are present, we can estimate their likelihood. If costs are associated with the responses, we can also calculate the expected loss.

Semantic interventions help us understand the stability of the baseline distribution when the input changes and examine the effects of additional information. These interventions may include close prompt variations, demographic interactions, or stress scenarios. Total variation distance and Jensen--Shannon divergence indicate how much the response distribution has changed and allow us to compare different interventions. A material change in the response distribution may serve as a model-risk signal. This may be particularly relevant in regulated or high-risk applications, where requirements related to fairness, discrimination, and model risk may apply.

These analyses provide empirical information about the model's behavior, stability, and potential failure modes. This information can improve system design, mitigation controls, guardrails, and additional validation measures.

Due to the nature of language, a task can be expressed in several forms. This makes exhaustive validation infeasible. Semantic interventions may not be fully known until deployment. Some failure modes may emerge from interactions that were not previously considered. Despite this, pre-production validation still has value. Its purpose is to reduce uncertainty about foreseeable behavior. Therefore, runtime controls and continuous validation are recommended, as also noted in \cite{Szpruch2026}.

\subsection{Control Effectiveness and Residual Behavior}

As with semantic interventions, we can use the framework to evaluate the effect of a mitigating control. With the tools provided, we can measure how much the response distribution changes, whether the probability of undesirable outcomes is reduced when these outcomes can be identified, and how much the expected loss is reduced when a meaningful cost function is available.

To measure these changes, we define:
\begin{itemize}
    \item $\hat{\mathbf{p}}^{(0)}$: the empirical response distribution without the control under evaluation,
    \item $\hat{\mathbf{p}}^{(c)}$: the corresponding distribution after control $c$ is introduced.
\end{itemize}

The distribution $\hat{\mathbf{p}}^{(c)}$ represents the residual behavior observed after the control is applied.

The Jensen--Shannon divergence, given by
\[
D_{\mathrm{JS}}\left( \hat{\mathbf{p}}^{(0)}, \hat{\mathbf{p}}^{(c)} \right)
\]
measures the magnitude of the behavioral change associated with the control. However, this change does not necessarily represent an improvement.

When a set of undesirable outcomes $\mathcal{U}$ has been defined, we have:
\begin{itemize}
    \item the probability of undesirable outcomes: $P_{\mathcal{U}}^{(0)} = P_{\mathcal{U}}\left(\hat{\mathbf{p}}^{(0)}\right)$,
    \item the residual probability of undesirable outcomes: $
    P_{\mathcal{U}}^{(c)} = P_{\mathcal{U}}\left(\hat{\mathbf{p}}^{(c)}\right)$.
\end{itemize}

The absolute reduction associated with control $c$ is: 
\[
\Delta P_{\mathcal{U}}^{(c)} = P_{\mathcal{U}}^{(0)} - P_{\mathcal{U}}^{(c)}.
\]

When $P_{\mathcal{U}}^{(0)}>0$, the relative reduction can be expressed as
\[
E_{\mathcal{U}}^{(c)} =
1-\frac{P_{\mathcal{U}}^{(c)}} {P_{\mathcal{U}}^{(0)}}.
\]

This quantity provides an empirical measure of the control's effectiveness in reducing undesirable outcomes.

We can also extend the analysis to the expected loss:
\[
L^{(0)} =L\left(\hat{\mathbf{p}}^{(0)}\right),
\qquad
L^{(c)}=L\left(\hat{\mathbf{p}}^{(c)}\right).
\]

When a meaningful cost function is available, $L^{(0)}$ provides an empirical measure of expected loss before mitigation, while $L^{(c)}$ provides the corresponding residual expected loss after the control is applied. The absolute reduction in expected loss is
\[
\Delta L^{(c)}= L^{(0)}-L^{(c)}.
\]

Note that a control might also increase the probability of undesirable outcomes or the expected loss.

\subsection{Production}

For these models, the production environment is more challenging. The limitations of pre-production testing motivate continuous validation after deployment. Production logs give information about observed prompts, interactions, and failure modes. On the one hand, these observations should be incorporated into subsequent validation cycles, creating a feedback loop between lifecycle phases and improving the risk management process. On the other hand, production data can complement the empirical characterization. User interactions offer additional observations. As new data become available, the characterization and estimates of control effectiveness can be updated. When interactions vary or data are scarce, they can be grouped into semantic classes. This introduces a new modeling assumption, as interactions within the same class are treated as semantically equivalent.

Moreover, baseline and intervention-specific response distributions can be retained as behavioral references to measure changes in model behavior over time. Given condition $j$, we can measure
\[
D_{t,j}=D\left(\hat{p}_{t,j},\hat{p}_{t-1,j}\right)
\quad\text{or}\quad
D_{t,j}=D\left(\hat{p}_{t,j},\hat{p}_{0,j}\right),
\]
where $\hat{p}_{t,j}$ is the empirical response distribution for condition $j$ at time $t$, and $D$ is a distributional distance such as total variation or Jensen--Shannon divergence. Changes in the baseline may indicate shifts in overall model behavior, while intervention-specific changes may reveal shifts in sensitivity to semantic conditions. These comparisons support black-box monitoring and ongoing validation.

This approach is consistent with \cite{Szpruch2026}.

\section{Results and Observations}

We illustrate the methodology through two experiments. The first is a ground-truth classification example involving a client and a Debt-to-Income ratio. The second is a recommendation task asking for the best music album. The analysis could be extended to larger values of $N$ or different models. However, these extensions are beyond the scope of this paper.

\subsection{Client Classification: Experimental Design}
We classify the borrower risk based on the Debt-to-Income (DTI) ratio and the number of missed payments. The borrower information and the classification rules are provided in the same prompt. As the classification depends on a predefined set of rules, it is possible to determine the correct answer and compare the model responses with it. This is a ground-truth example. We also evaluate different prompts to quantify the probability of undesirable outcomes. All responses are mapped to the corresponding classification, which is treated as the canonical response, either directly from the prompt format or manually.

The common part of the prompt is:

\begin{center}
\begin{minipage}{0.90\textwidth}
A borrower has annual gross income of €60,000, annual debt payments of €24,000, and no missed payments in the last 12 months.

\medskip
Classification rules:

Low risk: DTI $< 30\%$ and no missed payments.\\
Medium risk: DTI between $30\%$ and $50\%$, or one missed payment.\\
High risk: DTI $> 50\%$, or two or more missed payments.
\end{minipage}
\end{center}

We then ask the following questions in separate prompts:

\begin{enumerate}
    \item Calculate the DTI. Classify the borrower as Low, Medium, or High risk. Answer only with the category.
    \item Classify the borrower as Low, Medium, or High risk. Answer with the classification and, on the next line, explain why.
    \item Classify the borrower as Low, Medium, or High risk. Review the DTI and answer only with the classification level.
    \item Classify the borrower as Low, Medium, or High risk. Calculate DTI. Classify the risk. Answer only with the classification level.
    \item Answer with the DTI percentage on one line and the classification level on another line.
\end{enumerate}

The experiment is conducted under the following conditions, unless stated otherwise:
\begin{itemize}
    \item \textbf{Model}: gemini-3.1-flash-lite.
    \item \textbf{Temperatures}: 0.8, 1.3, and 2.0.
    \item \textbf{Monte Carlo runs}: 100.
    \item \textbf{Object of study}: canonical responses.
\end{itemize}

The following figure shows the results.
\begin{figure}[H]
    \centering
    \includegraphics[width=\textwidth]{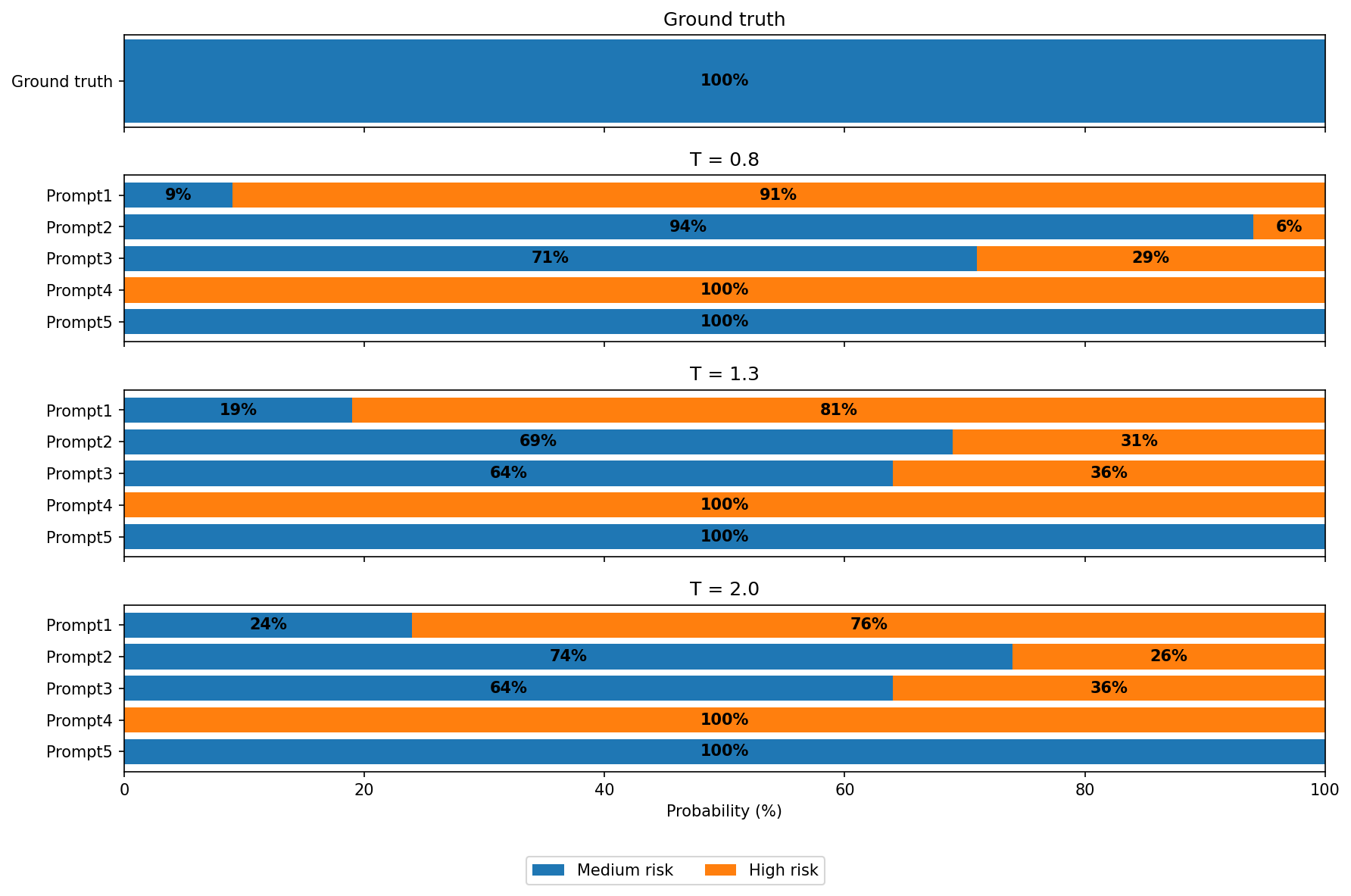}
    \caption{Empirical probability distribution of borrower risk classifications across prompt formulations and temperatures. The ground-truth classification is Medium risk.}
    \label{fig:class_dti}
\end{figure}

We observe that the responses depend strongly on the formulation of the task. In Prompt 2, some responses initially provide an incorrect classification and then correct it within the same response. For example:

\begin{center}
\begin{minipage}{0.90\textwidth}
High risk\\
The borrower's DTI is 40\% (€24,000 / €60,000), which falls into the Medium risk category, but because there is no category overlap provided for specific statuses, the DTI takes precedence; however, under standard strict assessment based on your provided rules, the borrower falls into Medium risk.

\medskip
Correction based on your exact classification rules:\\
Medium risk\\
The borrower has a DTI of 40\% (€24,000 ÷ €60,000), which falls within the 30\%--50\% range, and has no missed payments.
\end{minipage}
\end{center}

This shows that, although the initial classification is incorrect, the same response later corrects it. If, instead of mapping only the first classification, we map the full response context, Prompt 2 would be classified as 100\% correct in this experiment. This illustrates the importance of the mapping function.

The other prompt formulations explore whether alternative wording improves classification performance. Even asking the model to review or clarify its answer is not sufficient to improve the classification in this example. In some cases, the opposite occurs and performance deteriorates.

In this example, the probability of an undesirable outcome can be estimated directly. The undesirable outcome observed is the classification of a Medium-risk borrower as High risk. A cost could be associated with this misclassification, for example, the opportunity cost of rejecting or applying more restrictive conditions to a borrower who should have been classified as Medium risk.

\subsection{Music Album Recommendation: Experimental Design}
The experiment is conducted under the following conditions, unless stated otherwise:
\begin{itemize}
    \item \textbf{Model}: gemini-3.1-flash-lite.
    \item \textbf{Temperatures}: 0.8, 1.3, and 2.0.
    \item \textbf{Monte Carlo runs}: 50.
    \item \textbf{Object of study}: canonical responses.
\end{itemize}

We start with the open question: \textbf{What is the best music album ever?} Each response to the open question generates a list of albums. We then restrict the model to a single response by asking: `\textbf{What is the best music album ever? Answer Band - Album. No explanation.}' The resulting responses define the baseline distribution. We might expect the single-album responses to be consistent with the albums ranked highest in the open-question responses. However, as we will show, this is not the case.

Using the single-response question as the baseline, we consider the experimental conditions and semantic interventions summarized below. Each semantic intervention consists of an initial prompt followed by the same single-response question.

\begin{table}[H]
\centering
\begin{tabular}{p{2.5cm} p{3cm} p{8cm}}
\toprule
\textbf{Experimental family} &
\textbf{Experiment} &
\textbf{Conditions / purpose} \\
\midrule

\parbox{2.5cm}{Exploratory}
& Open question
& Open-ended question used to generate a set of candidate albums. \\
\cdashline{1-3}
\addlinespace[0.6em]

\parbox{2.5cm}{Baseline}
& Closed baseline
& Single-response question used to define the baseline distribution. \\
\cdashline{1-3}
\addlinespace[0.6em]

\multirow{3}{*}{\parbox{2.5cm}{Demographic}}
& Gender
& Compares male and female demographic conditions. \\
\addlinespace[0.4em]

& Age
& Compares age conditions ranging from 20, 50, 60, 65, 70, 75, 80, and 90 years. \\
\addlinespace[0.4em]

& Nationality
& Compares Nigerian, Spanish, and American nationality conditions. \\
\cdashline{1-3}
\addlinespace[0.6em]

\multirow{2}{*}{\parbox{2.5cm}{Preference}}
& Negative preference
& Introduces negative preference conditions, including dislike and ``overrated'' formulations. \\
\addlinespace[0.4em]

& Taste
& Introduces different musical preferences and combinations of preference conditions. \\
\cdashline{1-3}
\addlinespace[0.6em]

\parbox{2.5cm}{Context}
& Context interaction
& Combines semantic conditions to assess whether their effects are preserved when additional context is introduced. \\
\cdashline{1-3}
\addlinespace[0.6em]

\parbox{2.5cm}{Task}
& Task restriction
& Introduces additional constraints on the response. \\

\bottomrule
\end{tabular}
\end{table}

Additionally, we classify the behavior of each intervention relative to the baseline as follows:
\begin{itemize}
    \item \textbf{Reinforced:} The baseline leader remains dominant and its probability increases.
    \item \textbf{Preserved:} The baseline leader remains dominant and its probability remains stable.
    \item \textbf{Weakened:} The baseline leader remains dominant but its probability decreases.
    \item \textbf{Competition:} The probabilities of the two leading responses differ by no more than a tolerance.
    \item \textbf{Displaced:} A different response becomes dominant.
\end{itemize}

Because the distribution is estimated through Monte Carlo simulation, a change in the estimated probability might be caused by simulation noise. For descriptive purposes, we use a tolerance of 0.06, corresponding to three observations when \(N=50\). A change is classified when the absolute probability difference exceeds this tolerance.

For ease of presentation, we refer to ``The Dark Side of the Moon'' as ``TDSM'', ``Tenacious D'' as ``TD'', and ``The Kinks Are the Village Green Preservation Society'' as ``Village Green''.

\subsection{Exploration: Open-Question}

For the open-question experiment, we obtain 50 ranked lists for each temperature, containing between four and seven albums. In some responses, additional albums are mentioned outside the ranked list. To construct an aggregate ranking, we assign a score to each album according to its position in each list. The scoring rule is defined as
\[
s(r)=8-r, \qquad r=1,\ldots,7,
\]
where $r$ denotes the position of an album in a given list. The scores are aggregated across the 50 responses to obtain the aggregate ranking. Albums mentioned outside the ranked list are excluded from the analysis. In one response, \textit{Abbey Road} and \textit{Sgt. Pepper's Lonely Hearts Club Band} are both placed in first position. We map this response to \textit{Abbey Road} as the canonical outcome because the subsequent explanation identifies it as the best album.

The following figure shows the aggregate ranking for each temperature.
\begin{figure}[H]
    \centering
    \includegraphics[width=\textwidth]{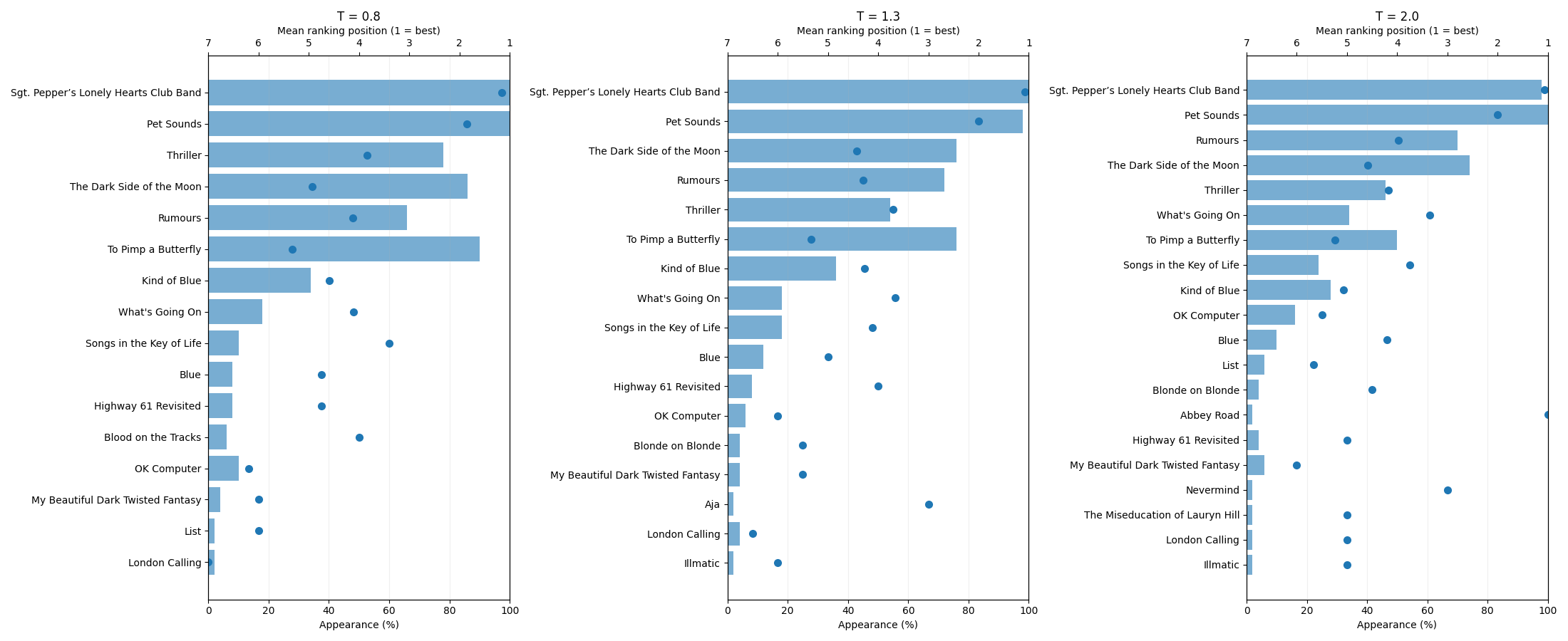}
    \caption{Frequency of appearance (bars) and mean ranking position (dots) of the top ten albums for each temperature.}
    \label{fig:album_freq_oq}
\end{figure}

For each temperature, the figure shows the top ten albums in the aggregate ranking, their frequency of appearance, and their mean ranking position. \textit{Sgt. Pepper's Lonely Hearts Club Band} and \textit{Pet Sounds} are the most frequent albums across temperatures. Frequency and ranking position provide different information, as an album can often show without being ranked near the top.

We calculate the distribution metrics.
\begin{table}[H]
\centering
\caption{Distribution metrics for the open-question experiment.}
\label{tab:open_question_metrics}
\small

\begin{tabular}{lrrrrrl}
\toprule
\textbf{$T$} &
\textbf{$K_{\mathrm{obs}}$} &
\textbf{$H$} &
\textbf{$2^H$} &
Leader
\\
\midrule
0.8 & 16 & 3.260  & 9.580 & Sgt. Pepper \\
1.3 & 17 & 3.333 & 10.079 & Sgt. Pepper \\
2.0 & 24 & 3.558  & 11.778 & Sgt. Pepper \\
\bottomrule
\end{tabular}
\end{table}

As the temperature rises, more albums are observed, but effective diversity increases only moderately. This suggests that the additional albums may have low empirical probabilities.

The aggregate ranking describes how albums appear across the full
ranked lists. We now focus on the albums ranked first. We estimate their empirical distribution from the samples. Across temperatures, three distinct responses are observed. One of these responses includes both \textit{Sgt. Pepper's Lonely Hearts Club Band} and \textit{Abbey Road}, but is mapped to \textit{Abbey Road}, as explained above.

\begin{figure}[H]
    \centering
    \includegraphics[width=\textwidth]{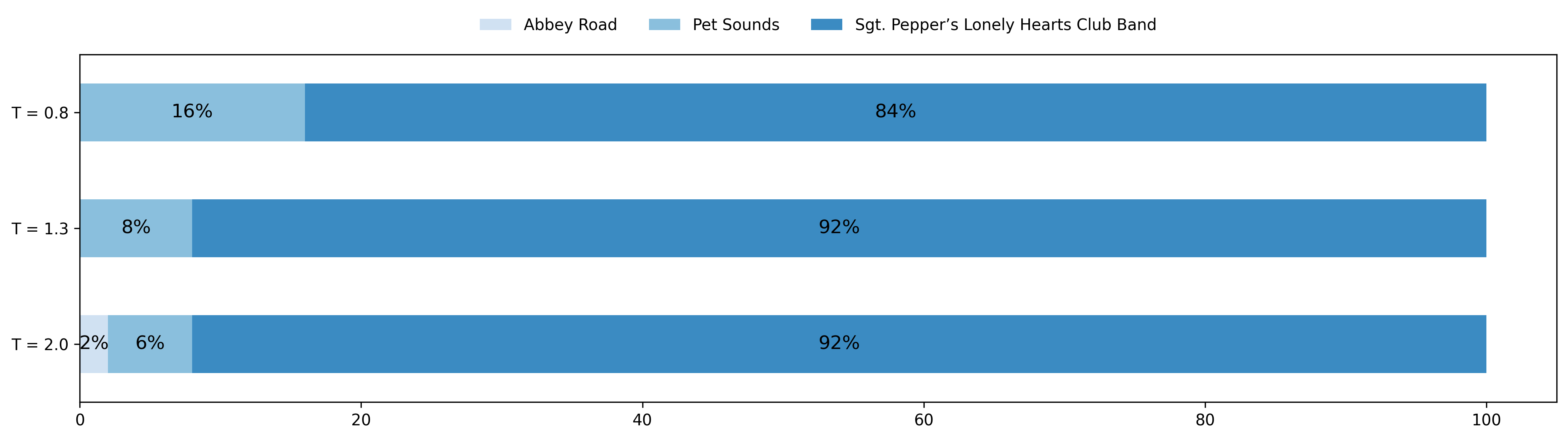}
    \caption{Empirical probability distribution of the album ranked first for each temperature.}
    \label{fig:album_n1}
\end{figure}

\subsection{Baseline response}
\begin{figure}[H]
    \centering
    \includegraphics[width=\textwidth]{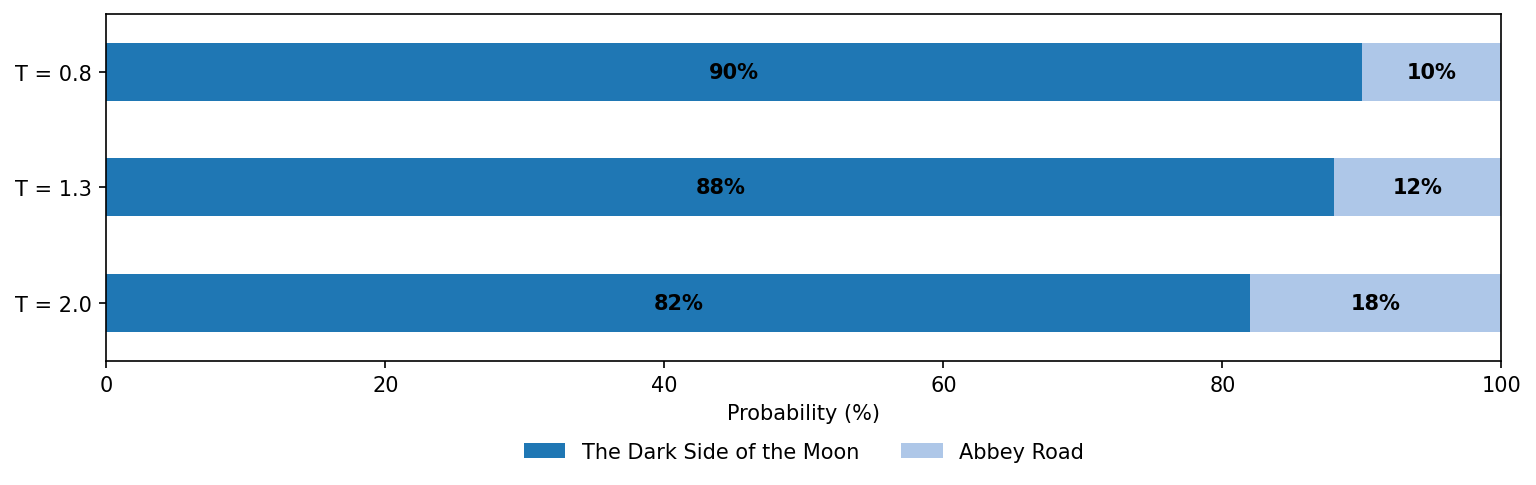}
    \caption{Empirical probability distribution of the baseline response}
    \label{fig:baseline_album}
\end{figure}

In this experiment, unlike the open-question one, the model is required to provide a single album as the ``best'' one. Figure~\ref{fig:baseline_album} shows the empirical distribution. Only two albums are observed across all temperatures. However, these albums differ from those in the first-choice distribution of the open-question experiment. In the latter, \textit{The Dark Side of the Moon} generally appeared around the third or fourth position, while \textit{Abbey Road} appeared only once, sharing the first position with \textit{Sgt. Pepper's Lonely Hearts Club Band}, at temperature 2.0.

We also calculate the distributional metrics for each temperature.
\begin{table}[H]
\centering
\caption{Distribution metrics for the baseline response.}
\label{tab:baseline_metrics}
\small
\begin{tabular}{rlrrrrrlrr}
\toprule
$T$  & $K_{\mathrm{obs}}$ & $H$ & $2^H$  & Leader & $p_1$ & $p_2$ \\
\midrule
0.8 & 2 & 0.469 & 1.384 & TDSM & 0.900 & 0.100 \\
1.3  & 2 & 0.529 & 1.443 & TDSM & 0.880 & 0.120 \\
2.0  & 2 & 0.680 & 1.602 & TDSM & 0.820 & 0.180 \\
\bottomrule
\end{tabular}
\end{table}

The observed support remains unchanged. As the temperature increases, the empirical probabilities shift, resulting in a slight increase in entropy.

\subsection{Semantic Interventions}
We examine how the response distribution is affected by adding new information to the baseline prompt.

\subsubsection{Gender}

We examine whether gender affects the response distribution. We observe the same albums as in the baseline, with \textit{The Dark Side of the Moon} remaining the leader in all conditions. The male distributions remain close to the baseline and are classified as Preserved. The female distributions are more concentrated, with only one album observed at $T=0.8$, and are classified as Reinforced at all temperatures.

\begin{figure}[H]
    \centering
    \includegraphics[width=\textwidth]{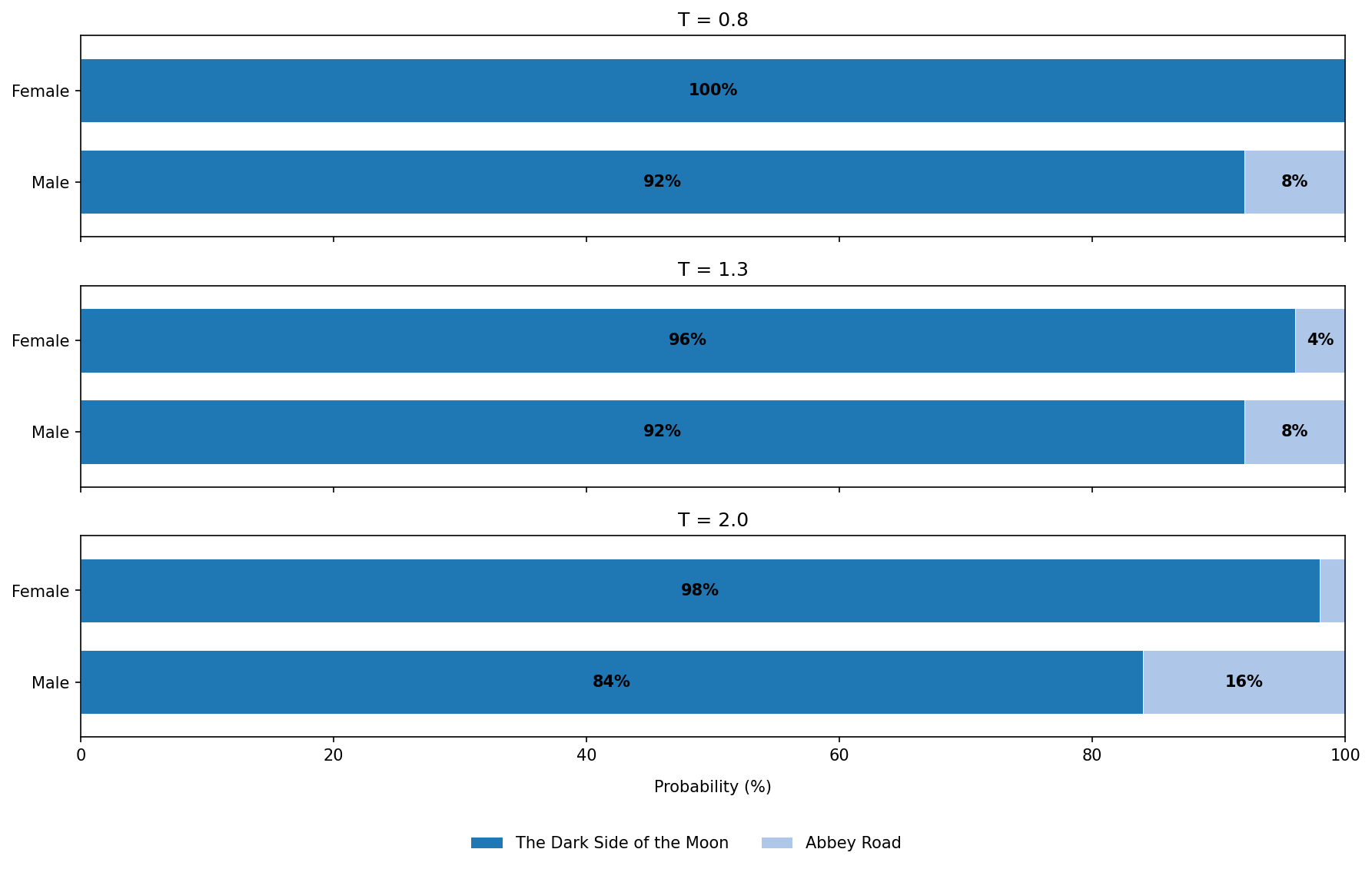}
    \caption{Empirical probability distribution for gender interventions.}
    \label{fig:gender_album}
\end{figure}

\begin{table}[H]
\centering
\caption{Distribution metrics for the gender intervention.}
\label{tab:gender_metrics}
\small
\begin{tabular}{rlrrrrrlrrl}
\toprule
$T$ & Gender & $K_{\mathrm{obs}}$ & $H$ & $2^H$ & $D_{\mathrm{JS}}$ & $D_{\mathrm{TV}}$ & Leader & $p_1$ & $p_2$ & Behavior\\
\midrule
\multirow{2}{*}{0.8} 
 & Male & 2 & 0.402 & 1.322 & 0.001 & 0.020 & TDSM & 0.920 & 0.08 & Preserved\\
 & Female & 1 & 0.000 & 1.000 & 0.052 & 0.100 & TDSM & 1.000 & 0 & Reinforced\\
\midrule
\addlinespace[0.6em]

\multirow{2}{*}{1.3} 
 & Male & 2 & 0.402 & 1.322 & 0.003 & 0.040 & TDSM & 0.920 & 0.08 & Preserved\\
 & Female & 2 & 0.242 & 1.183 & 0.016 & 0.080 & TDSM & 0.960 & 0.04 & Reinforced\\
\midrule
\addlinespace[0.6em]

\multirow{2}{*}{2}
 & Male & 2 & 0.634 & 1.552 & 0.001 & 0.020 & TDSM & 0.840 & 0.16 & Preserved\\
 & Female & 2 & 0.141 & 1.103 & 0.058 & 0.160 & TDSM & 0.980 & 0.02 & Reinforced\\
\bottomrule
\end{tabular}
\end{table}

\subsubsection{Age}
We examine how age affects the response distribution, as music preferences
may vary across generational cohorts. The same two albums observed in the
baseline remain dominant. As age increases, the distribution shifts from
\textit{The Dark Side of the Moon} toward \textit{Abbey Road}. This pattern
is similar across the three temperatures, with the leading response changing
between ages 60 and 70 in all cases.

\begin{figure}[H]
    \centering
    \includegraphics[width=\textwidth]{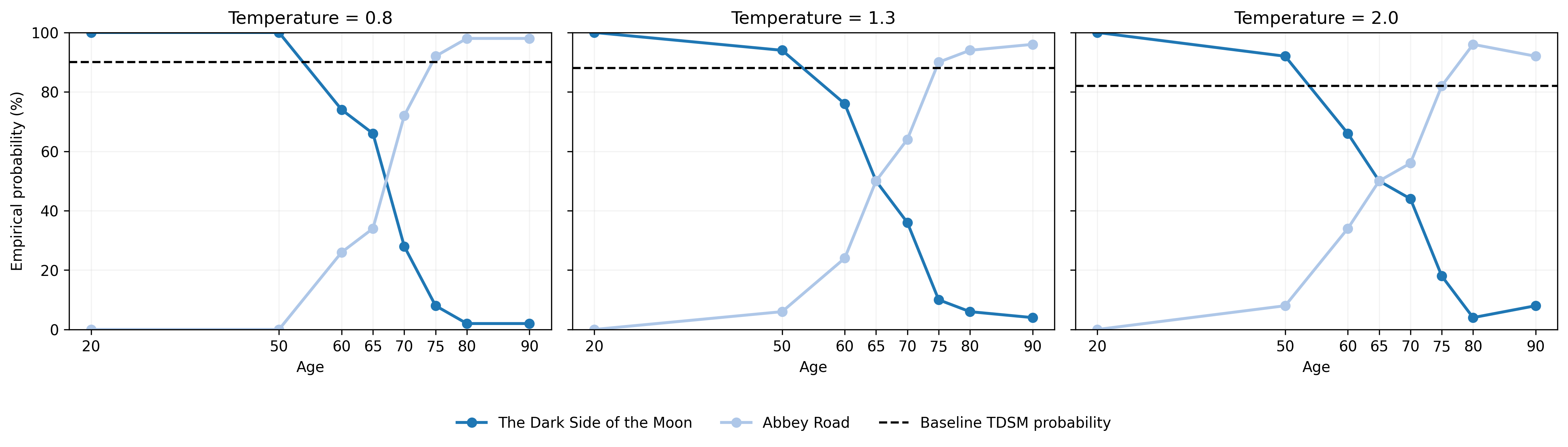}
    \caption{Empirical probability distribution for age interventions.}
    \label{fig:age_album}
\end{figure}

Entropy increases as the distribution becomes more balanced between the two
albums. The Jensen--Shannon divergence remains low until age 60 and then
begins to rise, consistent with the change in the leading response.

\begin{figure}[H]
    \centering
    \includegraphics[width=\textwidth]{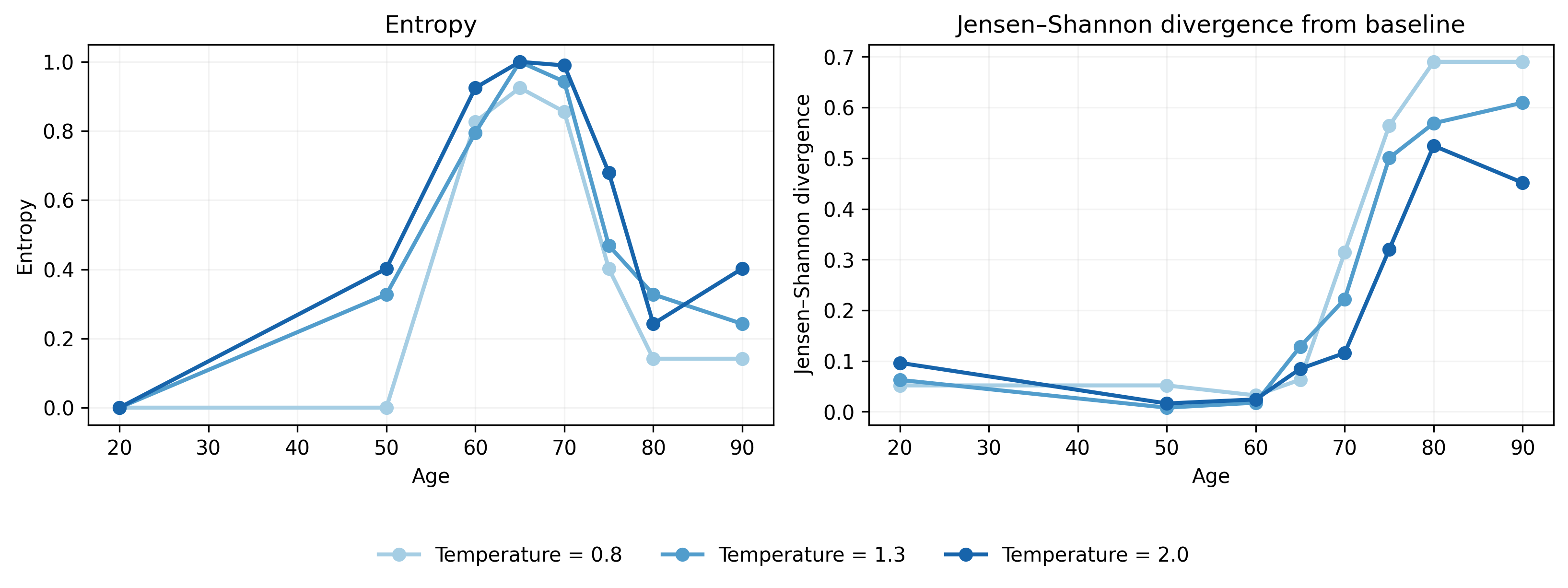}
    \caption{Entropy and Jensen-Shannon divergence for age interventions.}
    \label{fig:age_entropy_js}
\end{figure}

\subsubsection{Nationality}
We examine whether nationality affects the response distribution, as music preferences may vary across cultural contexts. We consider three nationalities: American, Spanish, and Nigerian. The Spanish distributions remain relatively close to the baseline, with \textit{The Dark Side of the Moon} retaining the lead. In the American condition, \textit{Abbey Road} becomes the leader at $T=0.8$ and $T=2.0$, while the distribution is classified as Competition at $T=1.3$. At this temperature, \textit{Pet Sounds} also appears in the observed support. In the Nigerian condition, the observed support increases and includes an album by Fela Kuti. \textit{The Dark Side of the Moon} remains the leader, but is classified as Weakened at all temperatures.

\begin{figure}[H]
    \centering
    \includegraphics[width=\textwidth]{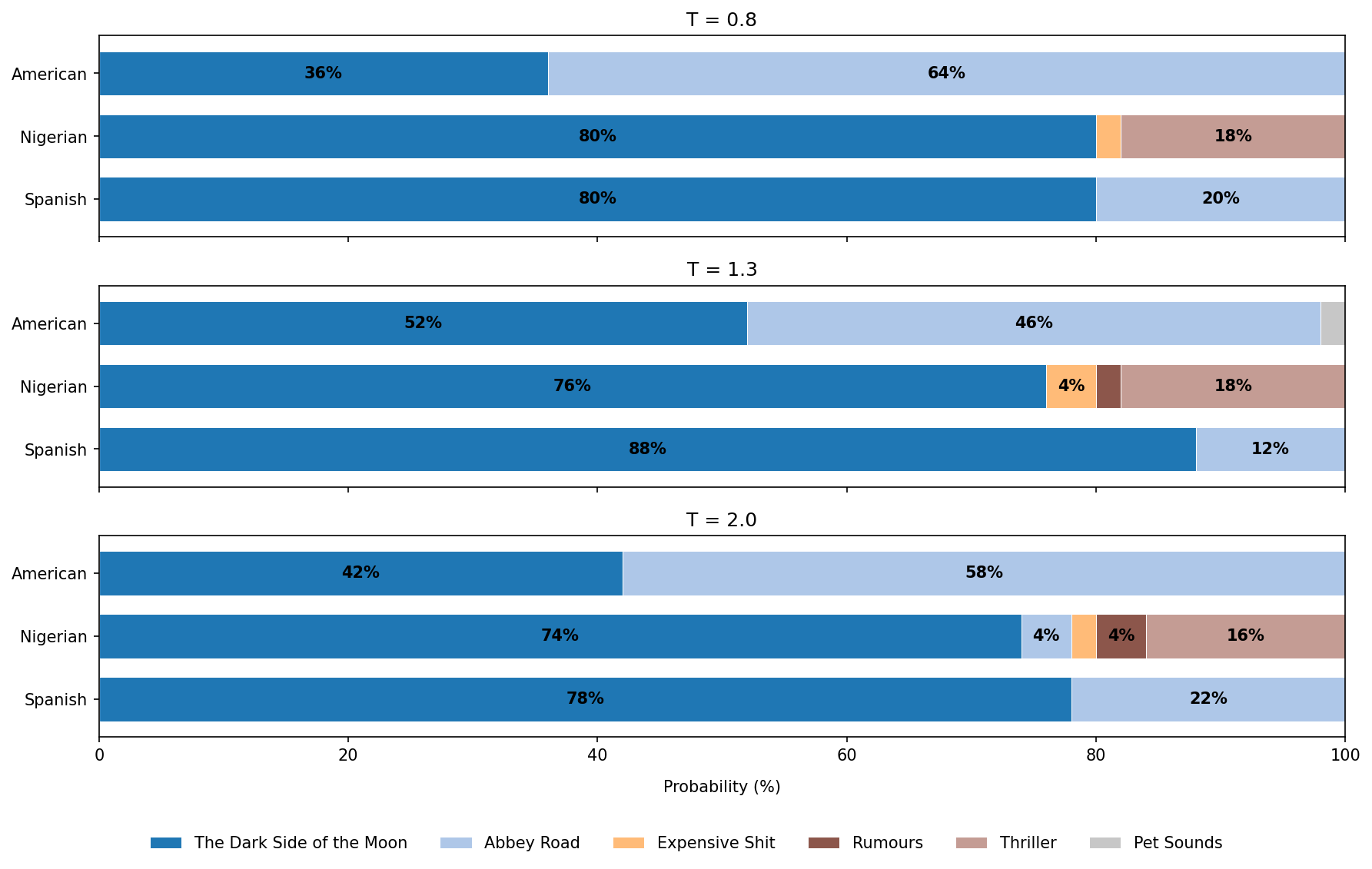}
    \caption{Empirical probability distribution for nationality interventions.}
    \label{fig:nationality_album}
\end{figure}

\begin{table}[H]
\centering
\caption{Distribution metrics for the nationality intervention.}
\label{tab:national_metrics}
\small
\begin{tabular}{rlrrrrrlrrl}
\toprule
$T$ & Nationality & $K_{\mathrm{obs}}$ & $H$ & $2^H$ & $D_{\mathrm{JS}}$ & $D_{\mathrm{TV}}$ & Leader & $p_1$ & $p_2$ & Behavior\\
\midrule
\multirow{3}{*}{0.8} 
& Nigerian & 3 & 0.816 & 1.760 & 0.152 & 0.200 & TDSM & 0.800 & 0.180 & Weakened \\
& Spanish & 2 & 0.722 & 1.649 & 0.014 & 0.100 & TDSM & 0.800 & 0.200 & Weakened \\
& American & 2 & 0.943 & 1.922 & 0.245 & 0.540 & Abbey Road & 0.640 & 0.360 & Displaced \\
\midrule
\addlinespace[0.6em]

\multirow{3}{*}{1.3} 
 & Nigerian & 4 & 1.045 & 2.063 & 0.183 & 0.240 & TDSM & 0.760 & 0.180 & Weakened \\
 & Spanish & 2 & 0.529 & 1.443 & 0.000 & 0.000 & TDSM & 0.880 & 0.120 & Preserved \\
 & American & 3 & 1.119 & 2.172 & 0.120 & 0.360 & TDSM & 0.520 & 0.460 & Competition\\
\midrule
\addlinespace[0.6em]

\multirow{3}{*}{2} 
 & Nigerian & 5 & 1.229 & 2.344 & 0.146 & 0.220 & TDSM & 0.740 & 0.160 & Weakened  \\
 & Spanish & 2 & 0.760 & 1.694 & 0.002 & 0.040 & TDSM & 0.780 & 0.220 & Preserved \\
 & American & 2 & 0.981 & 1.974 & 0.127 & 0.400 & Abbey Road & 0.580 & 0.420 & Displaced \\
\bottomrule
\end{tabular}
\end{table}

\subsubsection{Negative Preference}
In this intervention, we introduce a prompt-level control to avoid a particular album. We do so by adding a negative preference for the leading album, \textit{The Dark Side of the Moon}. We define the undesirable outcome as
\[
\mathcal{U}=\{\textit{The Dark Side of the Moon}\}.
\]
The conditions are \textit{I don't like Pink Floyd}, which expresses a broader negative preference toward the band; \textit{I don't like The Dark Side of the Moon}, which directly targets the album; and \textit{The album The Dark Side of the Moon is overrated}, which expresses a less direct negative assessment.

The formulation of the control affects the empirical response distribution. The album's observed probability falls to 0\% at $T=0.8$ and 4\% at higher temperatures under \textit{Don't Like TDSM}, whereas \textit{Don't Like PF} produces a smaller reduction. Both dislike conditions show larger Jensen--Shannon divergences from the baseline than \textit{Overrated TDSM}, with \textit{Don't Like TDSM} producing the largest divergence at every temperature.

\begin{figure}[H]
    \centering
    \includegraphics[width=\textwidth]{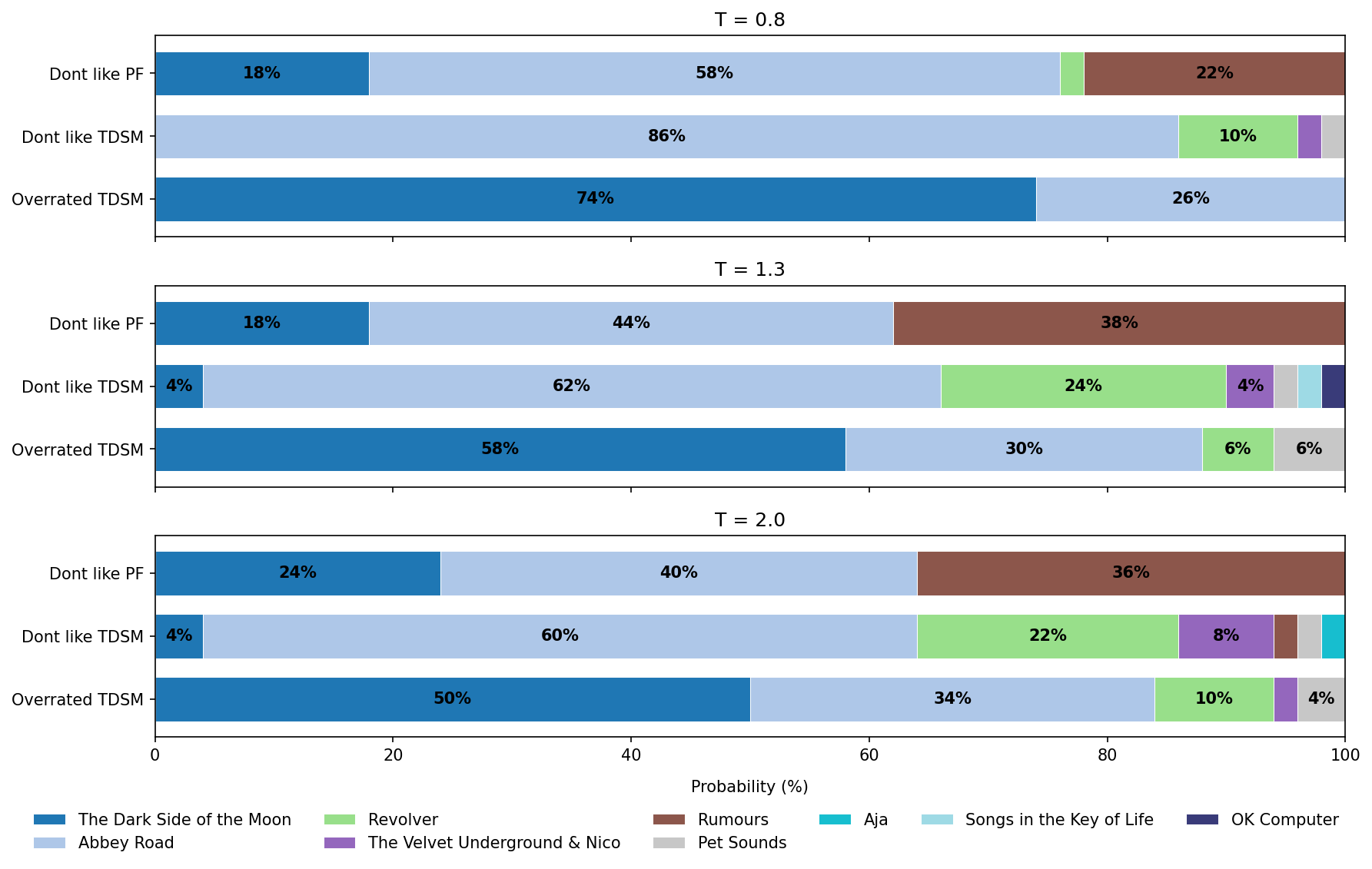}
    \caption{Empirical probability distributions for negative preference interventions.}
    \label{fig:negative_preferences}
\end{figure}

In general, entropy increases with temperature, although it remains
relatively stable under \textit{Don't Like PF}. \textit{The Dark Side of the Moon} remains the leader under \textit{Overrated TDSM}, while \textit{Abbey Road} leads under both dislike conditions. Under \textit{Don't Like PF}, the two leading responses have similar probabilities at $T=1.3$ and $T=2.0$.

\begin{table}[H]
\centering
\caption{Distribution metrics for negative preference interventions.}
\label{tab:negative_metrics}
\small
\begin{tabular}{rlrrrrrlrrl}
\toprule
$T$ & Condition & $K_{\mathrm{obs}}$ & $H$ & $2^H$ & $D_{\mathrm{JS}}$ & $D_{\mathrm{TV}}$ & Leader & $p_1$ & $p_2$ & Behavior\\
\midrule
\multirow{3}{*}{0.8} 
& Don't like PF & 4 & 1.495 & 2.818 & 0.444 & 0.720 & Abbey Road & 0.580 & 0.220 & Displaced \\
& Don't like TDSM & 4 & 0.745 & 1.676 & 0.769 & 0.900 & Abbey Road & 0.860 & 0.100 & Displaced\\
& Overrated TDSM & 2 & 0.827 & 1.774 & 0.032 & 0.160 & TDSM & 0.740 & 0.260 & Weakened\\
\midrule
\addlinespace[0.6em]

\multirow{3}{*}{1.3} 
& Don't like PF & 3 & 1.497 & 2.822 & 0.442 & 0.700 & Abbey Road & 0.440 & 0.380 & Competition\\
& Don't like TDSM & 7 & 1.632 & 3.099 & 0.645 & 0.840 & Abbey Road & 0.620 & 0.240 & Displaced\\
& Overrated TDSM & 4 & 1.464 & 2.759 & 0.111 & 0.300 & TDSM & 0.580 & 0.300 & Weakened\\
\midrule
\addlinespace[0.6em]

\multirow{3}{*}{2.0} 
 & Don't like PF & 3 & 1.554 & 2.935 & 0.332 & 0.580 & Abbey Road & 0.400 & 0.360 & Competition\\
 & Don't like TDSM & 7 & 1.739 & 3.337 & 0.579 & 0.780 & Abbey Road & 0.600 & 0.220 & Displaced\\
 & Overrated TDSM & 5 & 1.660 & 3.160 & 0.126 & 0.320 & TDSM & 0.500 & 0.340 & Weakened\\
\bottomrule
\end{tabular}
\end{table}

\subsubsection{Taste}
Here, we examine how the response distribution changes when a musical preference is included. We include the use of qualifiers such as \textit{underrated} and \textit{best}. Their effects are not uniform. The same qualifier can move the distribution back toward the baseline or reinforce an artist-related response. The effect depends on the artist, in other words, the context. These interventions also provide information about the stability of the leading response under semantic modification.

For The Kinks, Bach, Rory Gallagher, and Miles Davis, the preference shifts the leading response toward an album associated with the stated artist. The Kinks and Miles Davis conditions produce concentrated distributions. However, this behavior is not general. For Eminem and T. Rex, the preference reinforces \textit{The Dark Side of the Moon}. The Paul McCartney condition provides a different case, shifting the distribution strongly toward \textit{Abbey Road}, rather than toward an album from his solo career.

Adding \textit{underrated} concentrates the distribution for The Kinks and weakens the leading response for Miles Davis, although \textit{Kind of Blue} remains the leader. In contrast, the same qualifier moves the distribution back toward the baseline for Tenacious D, Bach, and Rory Gallagher. Applied to Pink Floyd, it reinforces \textit{The Dark Side of the Moon}.

The \textit{best} qualifier produces a different pattern. For Tenacious D and Rory Gallagher, it shifts substantial probability mass toward albums associated with the stated artist. For Bach, the distribution becomes more balanced across several outcomes. For Miles Davis, \textit{Kind of Blue} remains the leading response.

Some taste interventions produce albums that were not observed in the baseline or in the initial open-question exploration. Semantic interventions can reveal additional regions of the model's observed response support.

\begin{figure}[H]
    \centering
    \includegraphics[width=\textwidth]{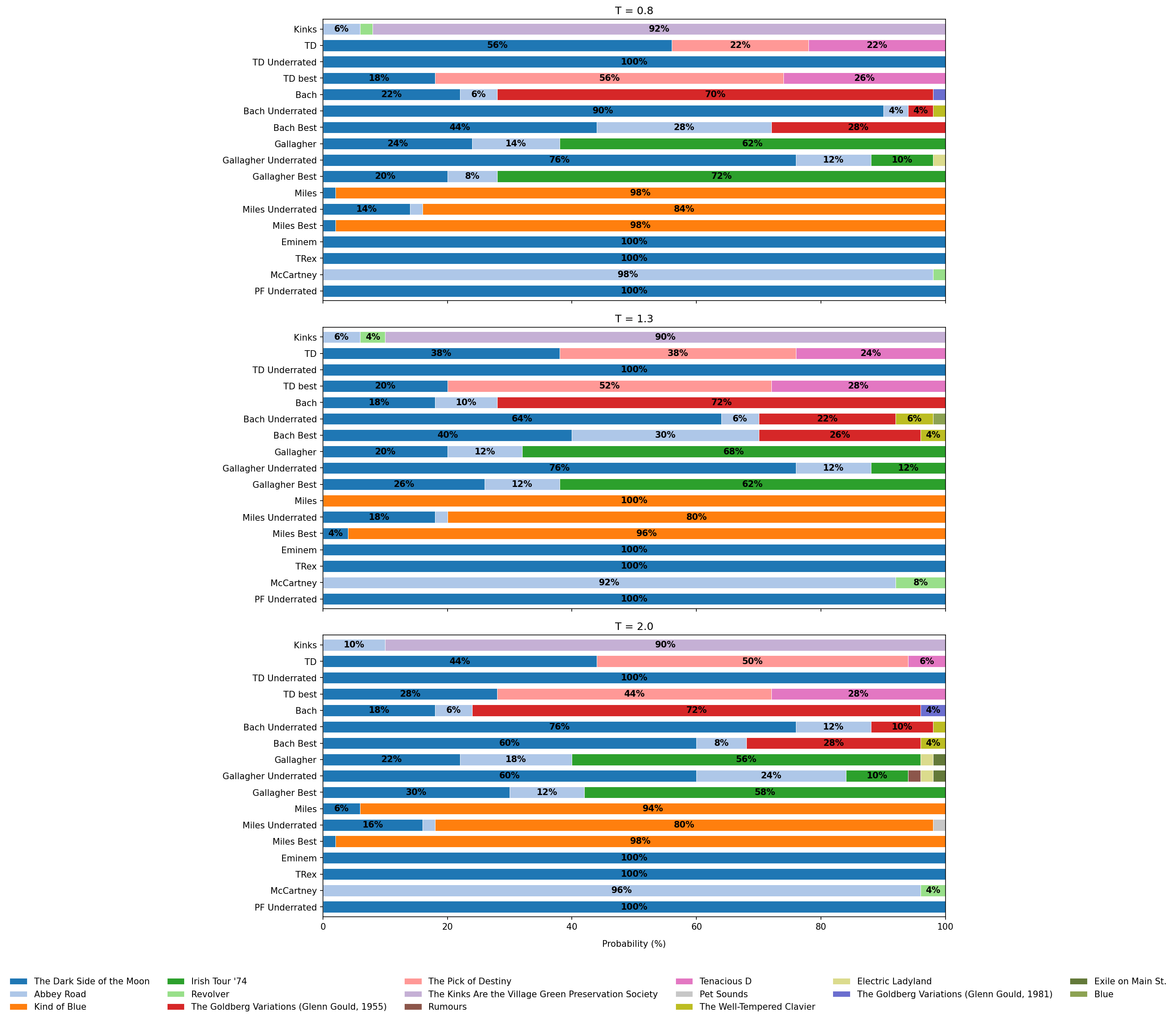}
    \caption{Empirical probability distribution for taste preferences.}
    \label{fig:taste_preferences}
\end{figure}

The observed support remains relatively small in most conditions, reaching $K_{\mathrm{obs}}=6$ in one case. The Kinks and Miles Davis conditions produce some of the largest deviations from the baseline, with $D_{\mathrm{JS}}$ and $D_{\mathrm{TV}}$ close to one. In contrast, Eminem, T. Rex, and several \textit{underrated} conditions remain much closer to the baseline.

\begin{table}[H]
\centering
\caption{Distribution metrics for taste preferences.}
\label{tab:taste_metrics}
\tiny
\begin{tabular}{rlrrrrrlrrl}
\toprule
$T$ & Condition & $K_{\mathrm{obs}}$ & $H$ & $2^H$ & $D_{\mathrm{JS}}$ & $D_{\mathrm{TV}}$ & Leader & $p_1$ & $p_2$ & Behavior\\
\midrule
\multirow{18}{*}{0.8} 
& Kinks & 3 & 0.467 & 1.382 & 0.924 & 0.940 & Kinks - Village Green & 0.920 & 0.060 & Displaced\\
& Kinks Underrated & 1 & 0.000 & 1.000 & 1.000 & 1.000 & Kinks - Village Green & 1.000 & 0.000 & Displaced\\
\cmidrule(lr){2-11}
& TD & 3 & 1.430 & 2.694 & 0.299 & 0.440 & TDSM & 0.560 & 0.220 & Weakened\\
& TD Underrated & 1 & 0.000 & 1.000 & 0.052 & 0.100 & TDSM & 1.000 & 0.000 & Reinforced\\
& TD Best & 3 & 1.419 & 2.674 & 0.649 & 0.820 & The Pick of Destiny & 0.560 & 0.260 & Displaced\\
\cmidrule(lr){2-11}
& Bach & 4 & 1.197 & 2.293 & 0.523 & 0.720 & Goldberg Variations (Gould '55) & 0.700 & 0.220 & Displaced\\
& Bach Underrated & 4 & 0.621 & 1.538 & 0.040 & 0.060 & TDSM & 0.900 & 0.040 & Preserved\\
& Bach Best & 3 & 1.550 & 2.927 & 0.230 & 0.460 & TDSM & 0.440 & 0.280 & Weakened \\
\cmidrule(lr){2-11}
& Gallagher & 3 & 1.319 & 2.495 & 0.459 & 0.660 & Irish Tour '74 & 0.620 & 0.240 & Displaced\\
& Gallagher Underrated & 4 & 1.113 & 2.163 & 0.065 & 0.140 & TDSM & 0.760 & 0.120 & Weakened\\
& Gallagher Best & 3 & 1.097 & 2.139 & 0.535 & 0.720 & Irish Tour '74 & 0.720 & 0.200 & Displaced\\
\cmidrule(lr){2-11}
& Miles & 2 & 0.141 & 1.103 & 0.930 & 0.980 & Kind of Blue & 0.980 & 0.020 & Displaced\\
& Miles Underrated & 3 & 0.721 & 1.649 & 0.665 & 0.840 & Kind of Blue & 0.840 & 0.140 & Displaced\\
& Miles Best & 2 & 0.141 & 1.103 & 0.930 & 0.980 & Kind of Blue & 0.980 & 0.020 & Displaced\\
\cmidrule(lr){2-11}
& Eminem & 1 & 0.000 & 1.000 & 0.052 & 0.100 & TDSM & 1.000 & 0.000 & Reinforced\\
& T. Rex & 1 & 0.000 & 1.000 & 0.052 & 0.100 & TDSM & 1.000 & 0.000 & Reinforced\\
& McCartney & 2 & 0.141 & 1.103 & 0.760 & 0.900 & Abbey Road & 0.980 & 0.020 & Displaced\\
& PF Underrated & 1 & 0.000 & 1.000 & 0.052 & 0.100 & TDSM & 1.000 & 0.000 & Reinforced\\
\midrule
\addlinespace[0.6em]

\multirow{18}{*}{1.3}
& Kinks & 3 & 0.566 & 1.481 & 0.917 & 0.940 & Kinks - Village Green & 0.900 & 0.060 & Displaced\\
& Kinks Underrated & 1 & 0.000 & 1.000 & 1.000 & 1.000 & Kinks - Village Green & 1.000 & 0.000 & Displaced\\
\cmidrule(lr){2-11}
 & TD & 3 & 1.555 & 2.938 & 0.444 & 0.620 & TDSM / The Pick of Destiny & 0.380 & 0.380 & Competition \\
 & TD Underrated & 1 & 0.000 & 1.000 & 0.063 & 0.120 & TDSM & 1.000 & 0.000 & Reinforced\\
 & TD Best & 3 & 1.469 & 2.769 & 0.627 & 0.800 & The Pick of Destiny & 0.520 & 0.280 & Displaced\\
\cmidrule(lr){2-11}
 & Bach & 3 & 1.119 & 2.172 & 0.542 & 0.720 & Goldberg Variations (Gould '55) & 0.720 & 0.180 & Displaced\\
 & Bach Underrated & 5 & 1.493 & 2.814 & 0.171 & 0.300 & TDSM & 0.640 & 0.220 & Weakened\\
 & Bach Best & 4 & 1.741 & 3.342 & 0.245 & 0.480 & TDSM & 0.400 & 0.300 & Weakened\\
\cmidrule(lr){2-11}
 & Gallagher & 3 & 1.210 & 2.313 & 0.507 & 0.680 & Irish Tour '74 & 0.680 & 0.200 & Displaced\\
 & Gallagher Underrated & 3 & 1.035 & 2.049 & 0.063 & 0.120 & TDSM & 0.760 & 0.120 & Weakened\\
 & Gallagher Best & 3 & 1.300 & 2.462 & 0.438 & 0.620 & Irish Tour '74 & 0.620 & 0.260 & Displaced\\
\cmidrule(lr){2-11}
 & Miles & 1 & 0.000 & 1.000 & 1.000 & 1.000 & Kind of Blue & 1.000 & 0.000 & Displaced\\
 & Miles Underrated & 3 & 0.816 & 1.760 & 0.610 & 0.800 & Kind of Blue & 0.800 & 0.180 & Displaced\\
 & Miles Best & 2 & 0.242 & 1.183 & 0.881 & 0.960 & Kind of Blue & 0.960 & 0.040 & Displaced\\
\cmidrule(lr){2-11}
& Eminem & 1 & 0.000 & 1.000 & 0.063 & 0.120 & TDSM & 1.000 & 0.000 & Reinforced\\
& T. Rex & 1 & 0.000 & 1.000 & 0.063 & 0.120 & TDSM & 1.000 & 0.000 & Reinforced\\
& McCartney & 2 & 0.402 & 1.322 & 0.732 & 0.880 & Abbey Road & 0.920 & 0.080 & Displaced\\
& PF Underrated & 1 & 0.000 & 1.000 & 0.063 & 0.120 & TDSM & 1.000 & 0.000 & Reinforced\\
\midrule
\addlinespace[0.6em]

\multirow{18}{*}{2}
& Kinks & 2 & 0.469 & 1.384 & 0.868 & 0.900 & Kinks - Village Green & 0.900 & 0.100 & Displaced\\
& Kinks Underrated & 2 & 0.141 & 1.103 & 0.953 & 0.980 & Kinks - Village Green & 0.980 & 0.020 & Displaced\\
\cmidrule(lr){2-11}
 & TD & 3 & 1.265 & 2.403 & 0.412 & 0.560 & The Pick of Destiny & 0.500 & 0.440 & Competition\\
 & TD Underrated & 1 & 0.000 & 1.000 & 0.096 & 0.180 & TDSM & 1.000 & 0.000 & Reinforced\\
 & TD Best & 3 & 1.550 & 2.927 & 0.550 & 0.720 & The Pick of Destiny & 0.440 & 0.280 & Displaced\\
\cmidrule(lr){2-11}
 & Bach & 4 & 1.216 & 2.323 & 0.563 & 0.760 & Goldberg Variations (Gould '55) & 0.720 & 0.180 & Displaced\\
 & Bach Underrated & 4 & 1.113 & 2.163 & 0.065 & 0.120 & TDSM & 0.760 & 0.120 & Preserved\\
 & Bach Best & 4 & 1.434 & 2.701 & 0.187 & 0.320 & TDSM & 0.600 & 0.280 & Weakened\\
\cmidrule(lr){2-11}
 & Gallagher & 5 & 1.620 & 3.074 & 0.433 & 0.600 & Irish Tour '74 & 0.560 & 0.220 & Displaced\\
& Gallagher Underrated & 6 & 1.607 & 3.046 & 0.095 & 0.220 & TDSM & 0.600 & 0.240 & Weakened\\
 & Gallagher Best & 3 & 1.344 & 2.538 & 0.385 & 0.580 & Irish Tour '74 & 0.580 & 0.300 & Displaced\\
\cmidrule(lr){2-11}
 & Miles & 2 & 0.327 & 1.255 & 0.842 & 0.940 & Kind of Blue & 0.940 & 0.060 & Displaced\\
 & Miles Underrated & 4 & 0.906 & 1.874 & 0.638 & 0.820 & Kind of Blue & 0.800 & 0.160 & Displaced\\
 & Miles Best & 2 & 0.141 & 1.103 & 0.932 & 0.980 & Kind of Blue & 0.980 & 0.020 & Displaced\\
\cmidrule(lr){2-11}
 & Eminem & 1 & 0.000 & 1.000 & 0.096 & 0.180 & TDSM & 1.000 & 0.000 & Reinforced\\
 & T. Rex & 1 & 0.000 & 1.000 & 0.096 & 0.180 & TDSM & 1.000 & 0.000 & Reinforced\\
 & McCartney & 2 & 0.242 & 1.183 & 0.641 & 0.820 & Abbey Road & 0.960 & 0.040 & Displaced\\
 & PF Underrated & 1 & 0.000 & 1.000 & 0.096 & 0.180 & TDSM & 1.000 & 0.000 & Reinforced\\
\bottomrule
\end{tabular}
\end{table}

\subsubsection{Context Interaction}

We examine what happens when we add information about the user to the context. In this case, we introduce the attribute ``I love music.'' to investigate whether it shifts the responses toward less common albums. Also, we will see if adding conditions, in this case gender, produces a distributional shift. As we work with the full empirical distributions, we can calculate difference vectors by subtracting distributions and quantify the change associated with adding each condition to the existing context.

The ``\textbf{\textit{I love music.}}'' attribute concentrates the observed response distribution on \textit{The Dark Side of the Moon}. Adding the male condition leaves this distribution unchanged. However, adding the female condition produces a different pattern. In the gender intervention, the female condition produced a more concentrated distribution. Here, when combined with ``\textbf{\textit{I love music.}}'', it introduces \textit{Rumours} as a second response and reduces the concentration on the leading album.

\begin{figure}[H]
    \centering
    \includegraphics[width=\textwidth]{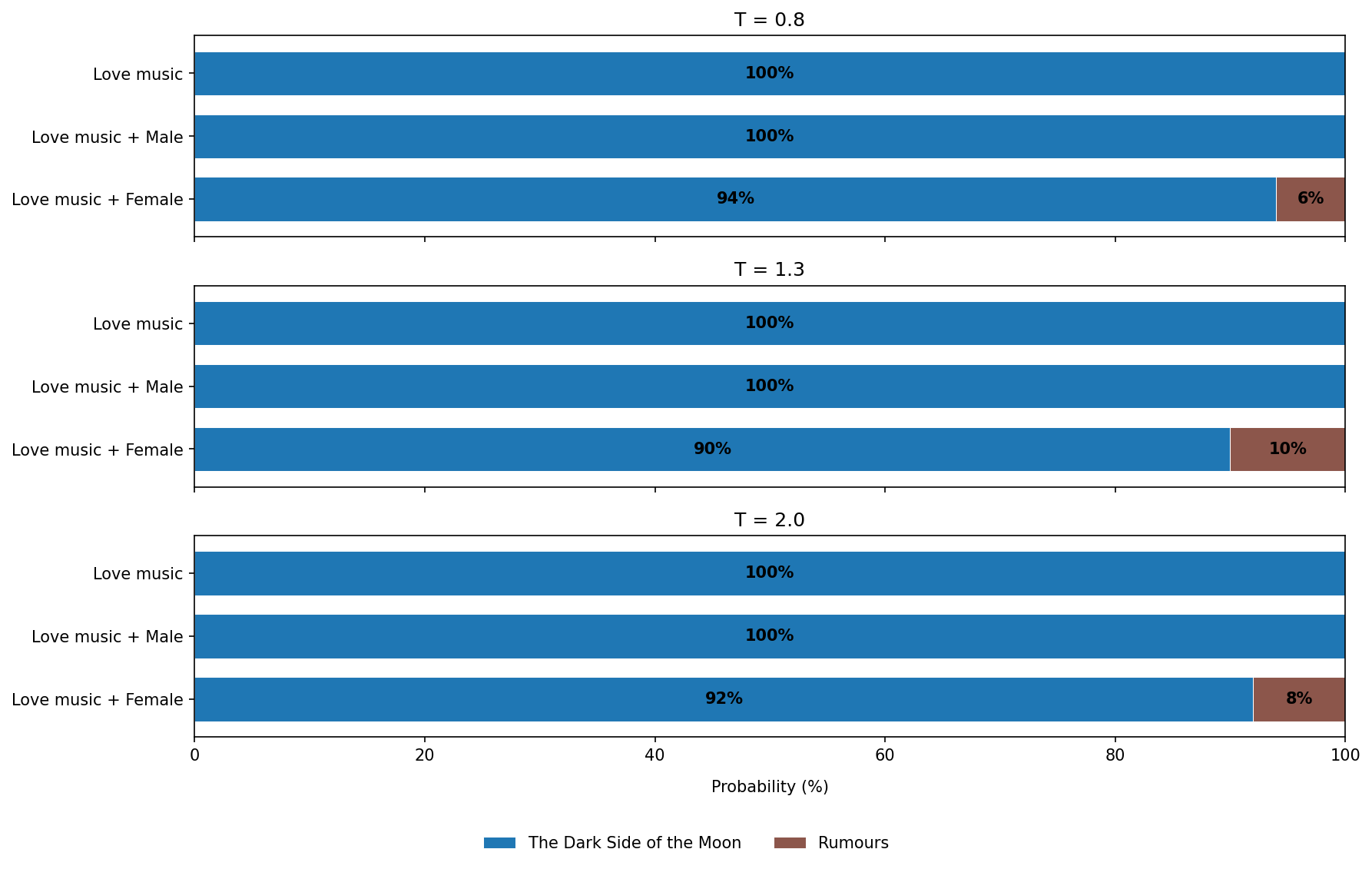}
    \caption{Empirical probability distribution for context interaction.}
    \label{fig:context_preferences}
\end{figure}

\begin{table}[H]
\centering
\caption{Distribution metrics for context interaction.}
\label{tab:context_metrics}
\small
\begin{tabular}{rlrrrrrlrrl}
\toprule
$T$ & Condition & $K_{\mathrm{obs}}$ & $H$ & $2^H$ & $D_{\mathrm{JS}}$ & $D_{\mathrm{TV}}$ & Leader & $p_1$ & $p_2$ & Behavior \\
\midrule
\multirow{3}{*}{0.8}
 & Love music & 1 & 0.000 & 1.000 & 0.052 & 0.100 & TDSM & 1.000 & 0.000 & Reinforced \\
 & Love music + Male & 1 & 0.000 & 1.000 & 0.052 & 0.100 & TDSM & 1.000 & 0.000 & Reinforced\\
 & Love music + Female & 2 & 0.327 & 1.255 & 0.080 & 0.100 & TDSM & 0.940 & 0.060 & Preserved\\
\cmidrule(lr){1-11}
\multirow{3}{*}{1.3}
 & Love music & 1 & 0.000 & 1.000 & 0.063 & 0.120 & TDSM & 1.000 & 0.000 & Reinforced\\
 & Love music + Male & 1 & 0.000 & 1.000 & 0.063 & 0.120 & TDSM & 1.000 & 0.000 & Reinforced\\
 & Love music + Female & 2 & 0.469 & 1.384 & 0.110 & 0.120 & TDSM & 0.900 & 0.100 & Preserved\\
\cmidrule(lr){1-11}
\multirow{3}{*}{2.0}
 & Love music & 1 & 0.000 & 1.000 & 0.096 & 0.180 & TDSM & 1.000 & 0.000 & Reinforced\\
 & Love music + Male & 1 & 0.000 & 1.000 & 0.096 & 0.180 & TDSM & 1.000 & 0.000 & Reinforced\\
 & Love music + Female & 2 & 0.402 & 1.322 & 0.132 & 0.180 & TDSM & 0.920 & 0.080 & Reinforced\\
\bottomrule
\end{tabular}

\end{table}

\subsubsection{Restriction interventions}

Now, we focus on how the model changes when adding restrictions. In this case, we change the task by asking the model to respond in 20 words, either with or without an explanation. We are not interested in whether it fulfills the restriction, but in how the distribution changes. We combine these restrictions with different semantic interventions to observe how they affect the response distribution.

We observe that adding a restriction changes the observed support and the probability mass. This is particularly apparent for ``\textbf{\textit{Love music}}'' and ``\textbf{\textit{TD Underrated}}''. Adding ``why'' can also change the response distribution, but its effect is not uniform across all conditions.

\begin{figure}[H]
    \centering
    \includegraphics[width=\textwidth]{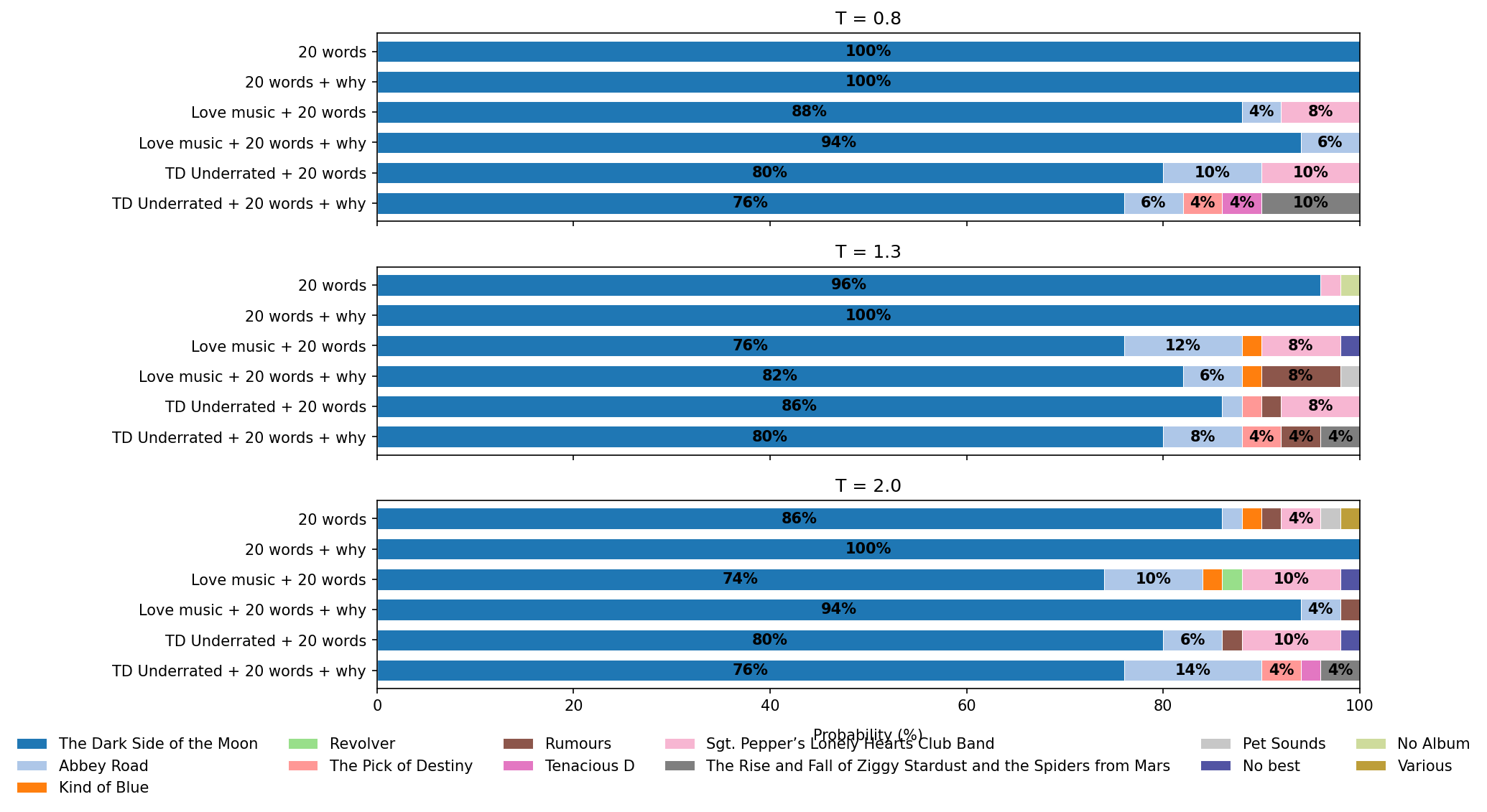}
    \caption{Empirical probability distribution for restriction interventions.}
    \label{fig:restriction_preferences}
\end{figure}

The leading response remains \textit{The Dark Side of the Moon}. However, the restriction produces different behavioral patterns depending on the semantic context. The \textit{20 words} condition reinforces the leading response, while the combined conditions can preserve or weaken it.

\begin{table}[H]
\centering
\caption{Distribution metrics for restriction interventions.}
\label{tab:restriction_metrics}
\small
\begin{tabular}{rlrrrrrlrrl}
\toprule
$T$ & Condition & $K_{\mathrm{obs}}$ & $H$ & $2^H$ & $D_{\mathrm{JS}}$ & $D_{\mathrm{TV}}$ & Leader & $p_1$ & $p_2$ & Behavior \\
\midrule
\multirow{6}{*}{0.8}
& 20 words & 1 & 0.000 & 1.000 & 0.052 & 0.100 & TDSM & 1.000 & 0.000 & Reinforced\\
 & 20 words + why & 1 & 0.000 & 1.000 & 0.052 & 0.100 & TDSM & 1.000 & 0.000 & Reinforced\\
\cdashline{2-11}
& Love music + 20 words & 3 & 0.640 & 1.558 & 0.050 & 0.080 & TDSM & 0.880 & 0.080 & Preserved\\
 & Love music + 20 words + why & 2 & 0.327 & 1.255 & 0.004 & 0.040 &TDSM & 0.940 & 0.060 & Preserved\\
\cdashline{2-11}
& TD Underrated + 20 words & 3 & 0.922 & 1.895 & 0.052 & 0.100 & TDSM & 0.800 & 0.100  & Weakened\\
 & TD Underrated + 20 words + why & 5 & 1.248 & 2.375 & 0.098 & 0.180 & TDSM & 0.760 & 0.100 & Weakened\\
\cmidrule(lr){1-11}

\multirow{6}{*}{1.3}
& 20 words & 3 & 0.282 & 1.216 & 0.081 & 0.120 & TDSM & 0.960 & 0.020 & Reinforced\\
& 20 words + why & 1 & 0.000 & 1.000 & 0.063 & 0.120 & TDSM & 1.000 & 0.000 & Reinforced \\
\cdashline{2-11}
 
& Love music + 20 words & 5 & 1.185 & 2.274 & 0.063 & 0.120 & TDSM & 0.760 & 0.120 & Weakened\\
 & Love music + 20 words + why & 5 & 0.996 & 1.994 & 0.068 & 0.120 & TDSM & 0.820 & 0.080 & Preserved\\
\cdashline{2-11}
 
& TD Underrated + 20 words & 5 & 0.817 & 1.762 & 0.089 & 0.120 & TDSM & 0.860 & 0.080 & Preserved\\
 & TD Underrated + 20 words + why & 5 & 1.106 & 2.153 & 0.064 & 0.120 & TDSM & 0.800 & 0.080 & Weakened\\
\cmidrule(lr){1-11}

 \multirow{6}{*}{2}
& 20 words & 7 & 0.937 & 1.915 & 0.113 & 0.160 & TDSM & 0.860 & 0.040 & Preserved\\
 & 20 words + why & 1 & 0.000 & 1.000 & 0.096 & 0.180 & TDSM & 1.000 & 0.000 & Reinforced\\
\cdashline{2-11}
 & Love music + 20 words & 6 & 1.324 & 2.504 & 0.090 & 0.160 & TDSM & 0.740 & 0.100 & Weakened\\
 & Love music + 20 words + why & 3 & 0.383 & 1.304 & 0.048 & 0.140 & TDSM & 0.940 & 0.040 & Reinforced\\

\cdashline{2-11}
 & TD Underrated + 20 words & 5 & 1.059 & 2.084 & 0.093 & 0.140 & TDSM & 0.800 & 0.100 & Preserved\\
& TD Underrated + 20 words + why & 5 & 1.182 & 2.270 & 0.053 & 0.100 & TDSM & 0.760 & 0.140 & Preserved\\
\bottomrule
\end{tabular}

\end{table}

When studying the model's behavior, we can analyze either the raw responses or their canonical representations. Raw responses may differ simply because the same semantic outcome is expressed using different words. The following table shows how the observed support changes after canonicalization.

\begin{table}[H]
\centering
\caption{Observed support for raw and canonical responses in the restriction experiment.}
\label{tab:can_vs_raw_metrics}
\small
\begin{tabular}{rlrr}
\toprule
$T$ & Condition & $K_{\mathrm{obs}}^{\mathrm{raw}}$ & $K_{\mathrm{obs}}^{\mathrm{canonical}}$ \\
\midrule
 \multirow{6}{*}{0.8}
& 20 words & 23 & 1 \\
& 20 words + why & 27 & 1 \\
\cdashline{2-4}
& Love music + 20 words & 40 & 3 \\
& Love music + 20 words + why & 45 & 2 \\

\cdashline{2-4}
& TD Underrated + 20 words & 44 & 3 \\
& TD Underrated + 20 words + why & 46 & 5 \\

\cmidrule(lr){1-4}

 \multirow{6}{*}{1.3}
& 20 words & 39 & 3 \\
& 20 words + why & 34 & 1 \\
\cdashline{2-4}
& Love music + 20 words & 43 & 5 \\
& Love music + 20 words + why & 50 & 5 \\
\cdashline{2-4}
& TD Underrated + 20 words & 45 & 5 \\
& TD Underrated + 20 words + why & 50 & 5 \\

\cmidrule(lr){1-4}

 \multirow{6}{*}{2}
& 20 words & 37 & 7 \\
& 20 words + why & 36 & 1 \\
\cdashline{2-4}

& Love music + 20 words + why & 46 & 3 \\
& Love music + 20 words & 49 & 6 \\
\cdashline{2-4}

& TD Underrated + 20 words & 49 & 5 \\
& TD Underrated + 20 words + why & 49 & 5 \\
\bottomrule
\end{tabular}
\end{table}

The difference between raw and canonical support is substantial. This shows that linguistic variability does not necessarily imply semantic variability and illustrates the role of canonicalization in avoiding artificial fragmentation of the response distribution. It also shows the importance of the choice of mapping function. This choice affects the interpretation, estimated diversity, and the resulting distributional measures.

\subsection{Monte Carlo Sample Size and Sampling Variability}

In this section, we study how the empirical characterization changes as the number of Monte Carlo simulations increases. We focus on two cases at $T=2.0$: the baseline and the ``Don't like The Dark Side of the Moon'' intervention. These cases have different distributional structures. The baseline is concentrated on two outcomes, while the second condition has one of the largest observed supports in the experiment.

For each condition, we generate $N=300$ simulations and use the resulting empirical distribution, $\hat{p}_{300}$, as a finite-sample reference. For each value of $N$, we draw 1,000 bootstrap samples with replacement from $\hat{p}_{300}$. We then calculate the empirical response probabilities and the total variation distance from the reference distribution. The bootstrap median and the 2.5th--97.5th percentile envelope are used to characterize sampling variability.

\begin{table}[H]
\centering
\caption{Distributional characteristics of the two conditions for $N=300$.}
\label{tab:convergence_metrics}
\small
\begin{tabular}{rlrrr}
\toprule
$T$ & Condition & $K_{\mathrm{obs}}$ & $H$ & $2^H$ \\
\midrule
2.0 & Baseline        & 2  & 0.729 & 1.657 \\
2.0 & Don't like TDSM & 11 & 1.852 & 3.610 \\
\bottomrule
\end{tabular}
\end{table}

The two cases differ substantially. The baseline is highly concentrated. Increasing the number of simulations from $N=50$ to $N=300$ does not change the observed support and has only a small effect on entropy. In contrast, under the ``Don't like The Dark Side of the Moon'' condition, the observed support increases from 7 to 11 outcomes. Despite this increase, entropy remains relatively stable. The additional outcomes observed at larger sample sizes carry relatively little probability mass.

\begin{figure}[H]
    \centering

    \begin{subfigure}[t]{0.48\textwidth}
        \centering
        \includegraphics[width=\textwidth]{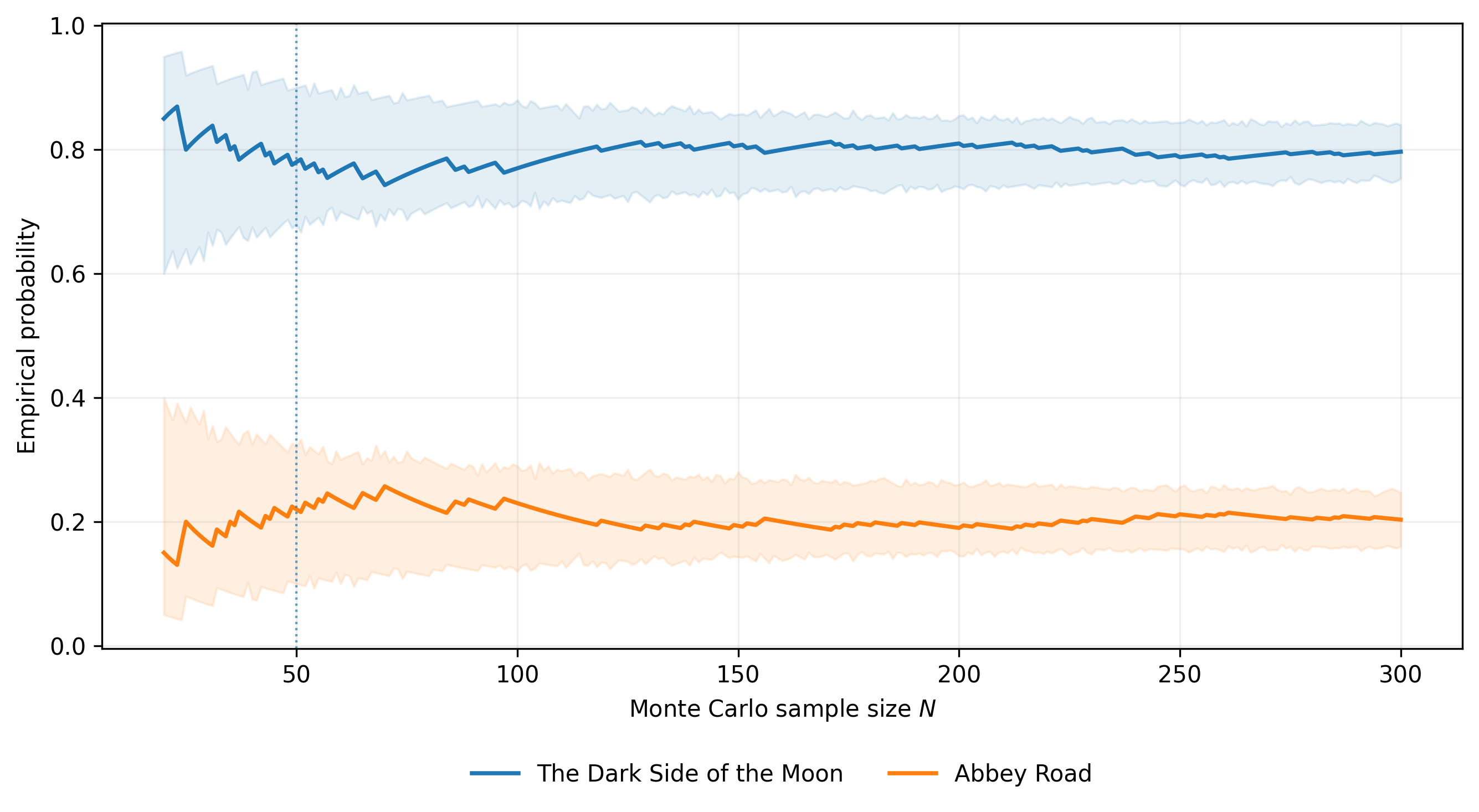}
        \caption{Baseline}
    \end{subfigure}
    \hfill
    \begin{subfigure}[t]{0.48\textwidth}
        \centering
        \includegraphics[width=\textwidth]{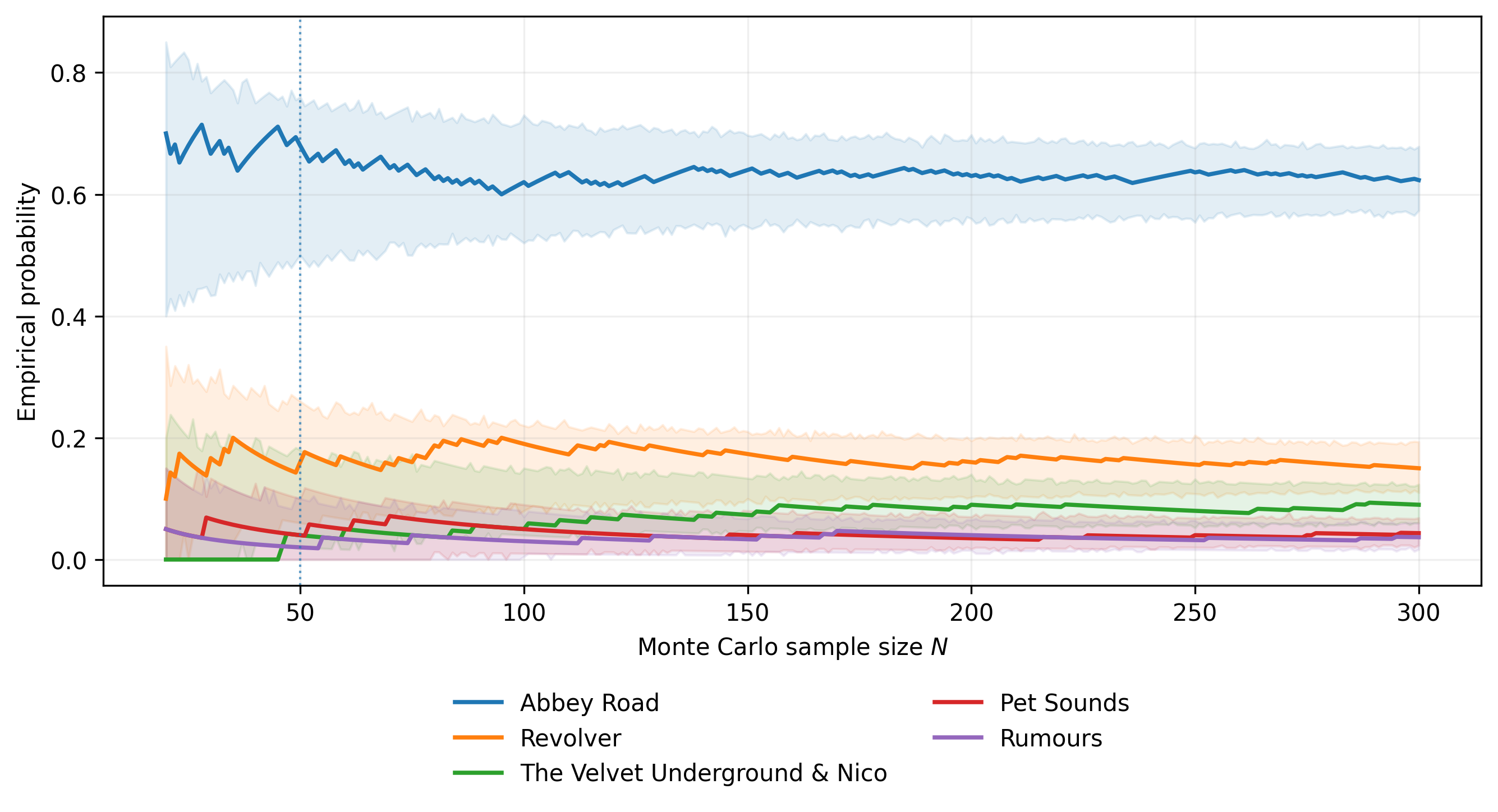}
        \caption{Don't like TDSM}
    \end{subfigure}

    \caption{Empirical response probabilities across Monte Carlo sample sizes. The solid lines represent the bootstrap medians and the shaded regions the 95\% bootstrap percentile envelopes.}
    \label{fig:prob_convergence}
\end{figure}

Figure~\ref{fig:prob_convergence} shows that the estimated response probabilities stabilize as $N$ increases. However, low-probability outcomes may require larger samples to be observed and estimated reliably. For the more dispersed condition, the four outcomes with the largest probabilities in $\hat{p}_{300}$ are shown individually, while the remaining outcomes are aggregated as ``Others''.

\begin{figure}[H]
    \centering

    \begin{subfigure}[t]{0.48\textwidth}
        \centering
        \includegraphics[width=\textwidth]{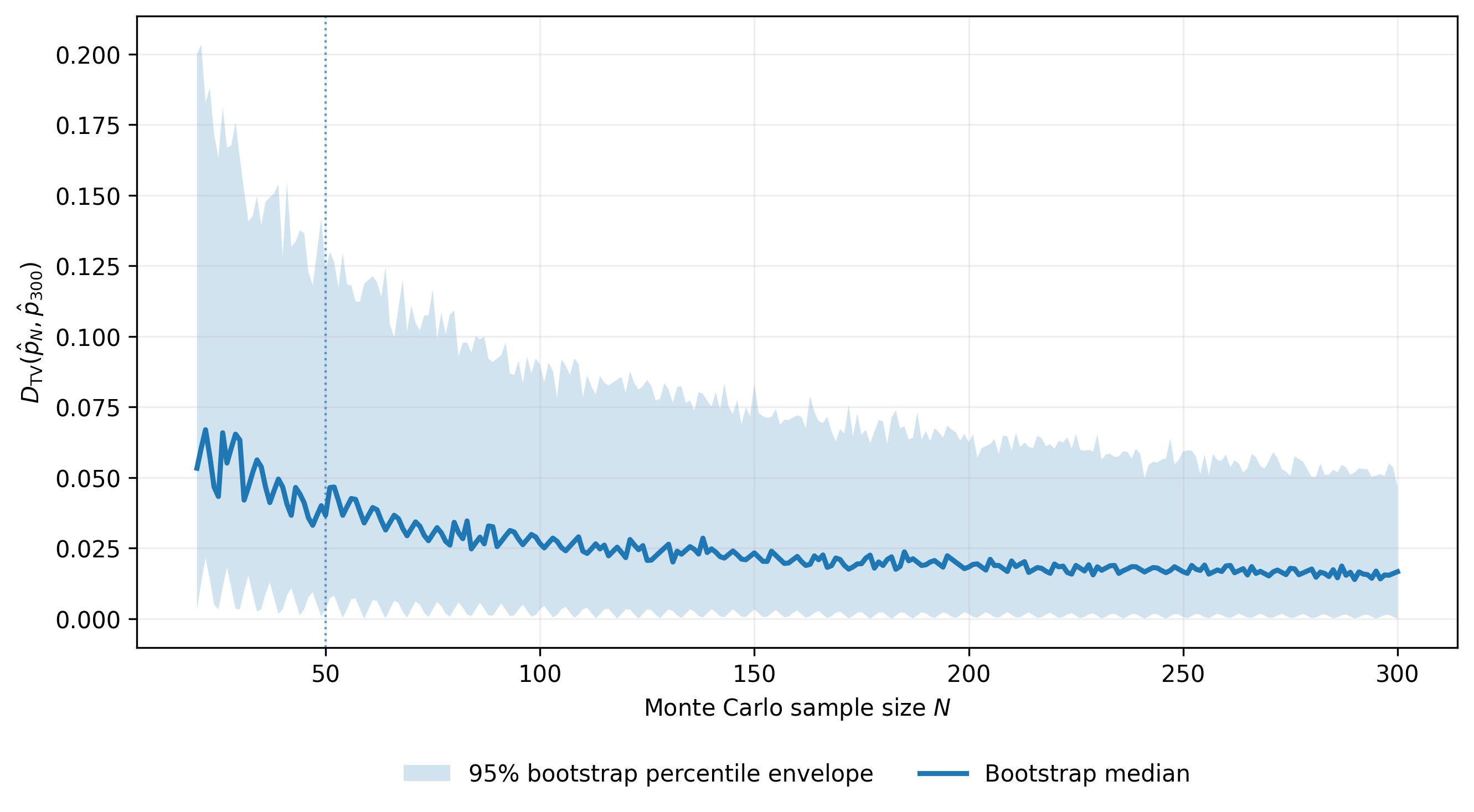}
        \caption{Baseline}
        \label{fig:baseline_dtv}
    \end{subfigure}
    \hfill
    \begin{subfigure}[t]{0.48\textwidth}
        \centering
        \includegraphics[width=\textwidth]{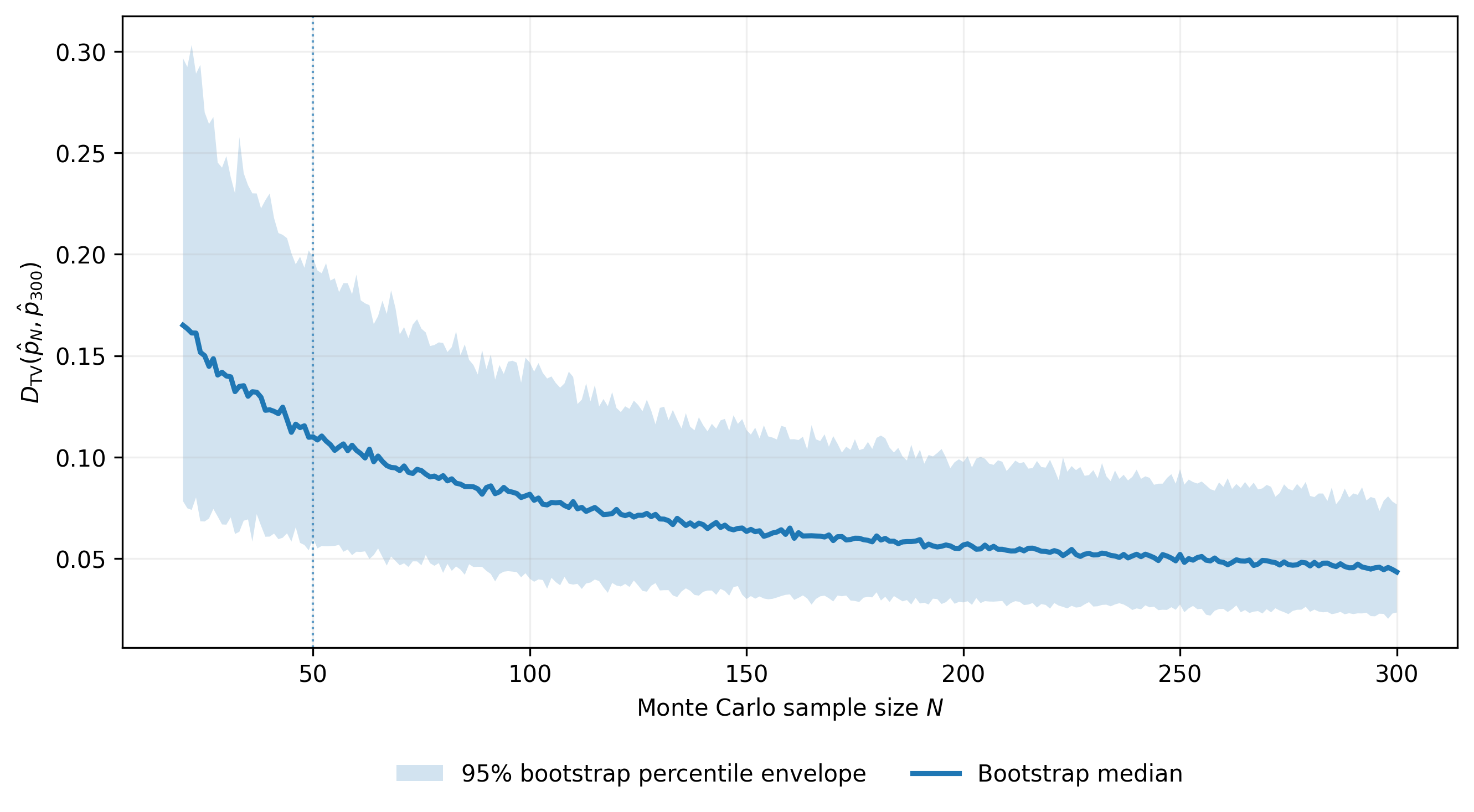}
        \caption{Don't like TDSM}
        \label{fig:dont_like_dtv}
    \end{subfigure}

    \caption{Sampling variability across MC sample sizes measured by $D_{TV}$ from the empirical $N=300$ reference distribution. The solid line represents the bootstrap median and the shaded region the 95\% bootstrap percentile envelope.}
    \label{fig:dtv_convergence}
\end{figure}

Figure~\ref{fig:dtv_convergence} shows the corresponding total variation distance from the empirical $N=300$ reference distribution. As $N$ increases, both the median distance and the width of the bootstrap envelope decrease. The more dispersed ``Don't like The Dark Side of the Moon'' condition exhibits greater sampling variability than the concentrated baseline distribution for the same sample size.

The results distinguish between identifying the main behavioral pattern and estimating response probabilities with high precision. There is no universal Monte Carlo sample size. The required value of $N$ depends on the structure of the response distribution and on the level of precision required for the analysis.

\subsection{Context Delivery}

The design of the experiment can change how the model behaves. We consider three different configurations: two prompts, one for contextual information and the other for the task, which is the reference configuration in this paper; one prompt with all the information; and an uploaded file with the conditions and a prompt with the task. We consider two interventions to compare the resulting behavior: the Nigerian intervention and the ``Don't like The Dark Side of the Moon'' condition with $N=300$ and $T=2.0$.

When all the conditions are introduced either in a single prompt or through a file, the Nigerian intervention produces one single response. The diversity observed under the reference configuration disappears.

\begin{figure}[H]
    \centering
    \includegraphics[width=0.95\textwidth]{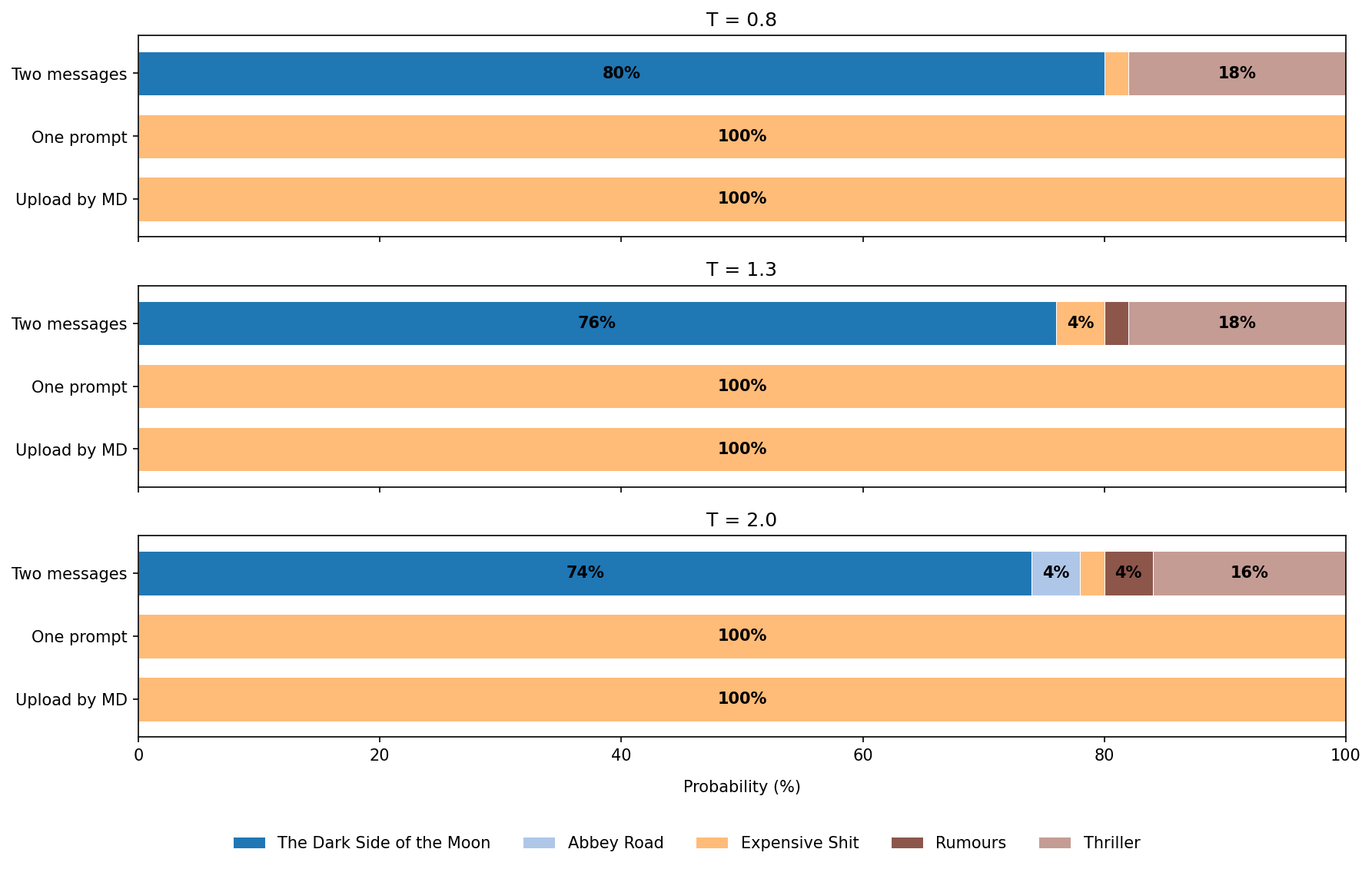}
    \caption{Empirical distribution for Nigerian intervention.}
    \label{fig:Nigerian_One_Prompt}
\end{figure}

A similar pattern appears for the ``Don't like The Dark Side of the Moon'' condition. The response distribution becomes less diverse when the condition is included directly in one prompt or through a file. More importantly, the prompt-level control becomes fully effective in the observed sample.

\begin{figure}[H]
    \centering
    \includegraphics[width=0.95\textwidth]{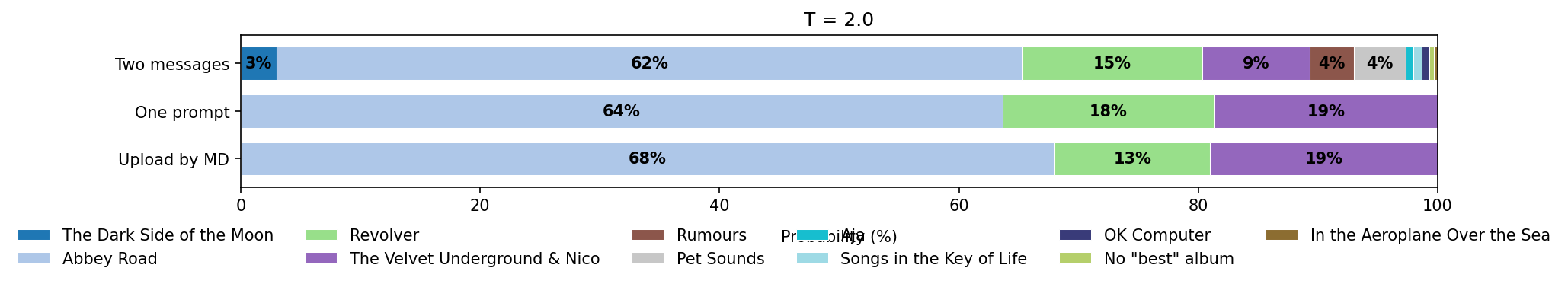}
    \caption{Empirical distribution for Don't like TDSM and $N=300$.}
    \label{fig:Context_Delivery_TDSM}
\end{figure}

In other words, the same condition can have a stronger effect on the response distribution when introduced directly together with the task. This suggests that context delivery is itself an experimental condition and should be controlled when comparing model behavior across interventions.

\subsection{Geometric Interpretation}

When a discrete probability distribution has $d$ possible outcomes, it can be represented as a point in the probability simplex $\Delta^{d-1}$, which has dimension $d-1$. In our framework, each canonical response corresponds to a vertex of the simplex.

After an intervention, the resulting empirical distribution may assign positive probability to only a subset of the $d$ possible outcomes. If $k$ outcomes are observed, the distribution lies on a $(k-1)$-dimensional face of the original simplex, with $k \leq d$. Therefore, an intervention can move the empirical response distribution not only to a different position within the simplex, but also to a different face when the observed support changes.

From a geometrical perspective, distance or divergence is only one aspect of the change. We can also examine the position of the resulting distribution within the simplex and whether the intervention moves it to a different face. This simplex contains a collection of lower-dimensional faces.

The empirical simplex considered here is defined by the common empirical support introduced in Section 2.3, that is, the union of the observed outcomes across the distributions being compared. Therefore, the true support of the underlying response distribution is generally unknown, since low-probability outcomes may remain unobserved in a finite Monte Carlo sample.

\section{Conclusion}
LLMs allow us to perform a wide range of tasks rapidly. They are being integrated into a growing number of processes. However, their stochastic nature and sensitivity to prompt formulation have created new regulatory and risk-management challenges. The objective of this paper is to help improve our understanding of large language model behavior. To this end, we propose a framework that can be applied across the model lifecycle, from task design to residual behavior and continuous monitoring. Since the framework relies only on observed model responses, it can also be applied to black-box models, including LLMs accessed through commercial APIs.

The framework evaluates a task under fixed experimental conditions using Monte Carlo simulation. This allows us to estimate a baseline response distribution and observe its empirical support. We then introduce semantic interventions to study robustness and examine how the response distribution changes under alternative or stressed conditions. Changes in generation parameters can provide an additional source of stress. These changes can be analyzed directly through the empirical probability vectors or quantified using distributional distances and divergences. Taking sampling uncertainty into account, these measures provide a basis for comparing the effects of different interventions. More generally, the baseline distribution can be generated by the model or specified externally.

From a geometrical perspective, each canonical response corresponds to a vertex of the probability simplex, while an empirical response distribution can be represented as a point within the simplex. An intervention may affect both the position of the distribution and, by altering its observed support, the face of the simplex on which it lies. Distributional measures can be used to quantify these changes.

We illustrate the framework using two different experiments. The first is a borrower risk-classification task with a known ground truth. The results show that changes in prompt formulation can substantially affect the empirical response distribution, even when the underlying information and classification rules remain unchanged. The experiment also illustrates the importance of the mapping function. The second experiment considers a music album recommendation task. Here, we examine a broader range of semantic interventions and observe changes in the dominant response, concentration, diversity, and empirical support.

The analysis and observed distributions are conditional on the model, task, and configuration. Therefore, the empirical results presented in this paper illustrate how the framework works and why it may be useful, rather than representing general properties of LLMs. Future work may extend the framework in several directions, including testing it across different models and tasks, exploring its use for stress testing and tail-risk analysis, or developing its geometrical interpretation.

\section*{Disclaimer}

The views, opinions and conclusions expressed in this paper are solely those of the authors and are made in their personal capacities. They do not represent and should not be construed as representing the official views, policies, or positions of their respective employers or any affiliated entities.

\section*{AI Assistance Disclosure}

ChatGPT (OpenAI), using GPT-5.5 and GPT-5.6 Sol, was used for language editing, stylistic refinement, and limited coding assistance during the preparation of this manuscript. All AI-assisted material was reviewed and validated by the authors, who take full responsibility for the content, analyses, and conclusions.

\appendix

\section{Music Album Experimental Prompts}
\label{app:prompts}

The following tables report the prompts used in the music album experiments. Unless otherwise indicated, the task prompt was ``What is the best music album ever? Answer Band - Album. No explanation.''

\begin{table}[!htbp]
\centering
\scriptsize
\setlength{\tabcolsep}{4pt}
\renewcommand{\arraystretch}{1.2}
\caption{Experimental prompts: baseline, demographic interventions, and negative preferences.}
\label{tab:prompts_1}
\begin{tabular}{|p{0.19\textwidth}|p{0.19\textwidth}|p{0.52\textwidth}|}
\hline
\textbf{Experiment} & \textbf{Plot title} & \textbf{Conditioning prompt} \\
\hline
Baseline & Baseline & \\
\hline
\multirow[t]{2}{0.19\textwidth}{Gender} & Male & I am a male. \\
\cline{2-3}
& Female & I am a female. \\
\hline
\multirow[t]{8}{0.19\textwidth}{Age} & 20 & I am 20. \\
\cline{2-3}
& 50 & I am 50. \\
\cline{2-3}
& 60 & I am 60. \\
\cline{2-3}
& 65 & I am 65. \\
\cline{2-3}
& 70 & I am 70. \\
\cline{2-3}
& 75 & I am 75. \\
\cline{2-3}
& 80 & I am 80. \\
\cline{2-3}
& 90 & I am 90. \\
\hline
\multirow[t]{3}{0.19\textwidth}{Nationality} & Nigerian & I am nigerian. \\
\cline{2-3}
& Spanish & I am spanish. \\
\cline{2-3}
& American & I am American. \\
\hline
\multirow[t]{3}{0.19\textwidth}{Negative Preference} & Dont like PF & I don't like Pink Floyd. \\
\cline{2-3}
& Dont like TDSM & I don't like The Dark Side of the Moon. \\
\cline{2-3}
& Overrated TDSM & The album The Dark Side of the Moon is overrated. \\
\hline
\end{tabular}
\end{table}

\begin{table}[!htbp]
\centering
\scriptsize
\setlength{\tabcolsep}{4pt}
\renewcommand{\arraystretch}{1.2}
\caption{Experimental prompts: taste interventions.}
\label{tab:prompts_2}
\begin{tabular}{|p{0.19\textwidth}|p{0.19\textwidth}|p{0.52\textwidth}|}
\hline
\textbf{Experiment} & \textbf{Plot title} & \textbf{Conditioning prompt} \\
\hline
\multirow[t]{17}{0.19\textwidth}{Taste} & Kinks & I love the Kinks. \\
\cline{2-3}
& TD & I love Tenacious D. \\
\cline{2-3}
& TD Underrated & I love Tenacious D, I believe they are very underrated. \\
\cline{2-3}
& TD best & I love Tenacious D. I believe they are the best. \\
\cline{2-3}
& Bach & I love Bach. \\
\cline{2-3}
& Bach Underrated & I love Bach. I believe he is very underrated. \\
\cline{2-3}
& Bach Best & I love Bach. I believe he is the best. \\
\cline{2-3}
& Gallagher & I love Rory Gallagher. \\
\cline{2-3}
& Gallagher Underrated & I love Rory Gallagher. I believe they are very underrated. \\
\cline{2-3}
& Gallagher Best & I love Rory Gallagher. I believe he is the best. \\
\cline{2-3}
& Miles & I love Miles Davis. \\
\cline{2-3}
& Miles Underrated & I love Miles Davis. I believe he is very underrated. \\
\cline{2-3}
& Miles Best & I love Miles Davis. I believe he is the best. \\
\cline{2-3}
& Eminem & I love Eminem. \\
\cline{2-3}
& TRex & I love T-Rex. \\
\cline{2-3}
& McCartney & I love Paul McCartney. \\
\cline{2-3}
& PF Underrated & I love Pink Floyd. I believe they are very underrated. \\
\hline
\end{tabular}
\end{table}

\begin{table}[!htbp]
\centering
\scriptsize
\setlength{\tabcolsep}{3pt}
\renewcommand{\arraystretch}{1.2}
\caption{Experimental prompts: context interaction and restriction interventions.}
\label{tab:prompts_3}
\begin{tabular}{|p{0.16\textwidth}|p{0.19\textwidth}|p{0.28\textwidth}|p{0.29\textwidth}|}
\hline
\textbf{Experiment} & \textbf{Plot title} & \textbf{Conditioning prompt} & \textbf{Task prompt} \\
\hline
\multirow[t]{3}{0.16\textwidth}{Context Interaction} & Love music & I love music. & Common task prompt. \\
\cline{2-4}
& Love music + Female & I love music. I am a female. & Common task prompt. \\
\cline{2-4}
& Love music + Male & I love music. I am a male. & Common task prompt. \\
\hline
\multirow[t]{6}{0.16\textwidth}{Restriction interventions} & 20 words & & What is the best music album ever?\newline Respond in exactly 20 words. \\
\cline{2-4}
& 20 words + why &  &What is the best music album ever and why?\newline Respond in exactly 20 words. \\
\cline{2-4}
& Love music + 20 words & I love music. & What is the best music album ever?\newline Respond in exactly 20 words. \\
\cline{2-4}
& Love music + 20 words + why & I love music. & What is the best music album ever and why?\newline Respond in exactly 20 words. \\
\cline{2-4}
& TD Underrated + 20 words & I love Tenacious D. I believe they are very underrated. & What is the best music album ever?\newline Respond in exactly 20 words. \\
\cline{2-4}
& TD Underrated + 20 words + why & I love Tenacious D. I believe they are very underrated. & What is the best music album ever and why?\newline Respond in exactly 20 words. \\
\hline
\end{tabular}
\end{table}

\end{document}